\documentclass[iop]{emulateapj}

\usepackage{apjfonts}
\usepackage{CJK}
\usepackage{color}
\usepackage{soul}
\usepackage[breaklinks,colorlinks,citecolor=blue,linkcolor=magenta]{hyperref}

\shorttitle{apogee chemical tagging constraint on cluster mass}
\shortauthors{Ting, Conroy, \& Rix}

\begin{document}

\begin{CJK*}{UTF8}{gbsn}
\title{APOGEE chemical tagging constraint on the maximum star cluster mass\\ in the $\alpha$-enhanced Galactic disk}
\author{Yuan-Sen Ting (丁源森) \altaffilmark{1}, Charlie Conroy\altaffilmark{1}, Hans-Walter Rix\altaffilmark{2}}
\altaffiltext{1}{Harvard--Smithsonian Center for Astrophysics, 60 Garden Street, Cambridge, MA 02138, USA}
\altaffiltext{2}{Max Planck Institute for Astronomy, K\"onigstuhl 17, D-69117 Heidelberg, Germany}

\slugcomment{Submitted to ApJ}

%
%
%
%
%
%
\begin{abstract}
Stars born from the same molecular cloud should be nearly homogeneous in their element abundances. The concept of chemical tagging is to identify members of disrupted clusters by their clustering in element abundance space. Chemical tagging requires large samples of stars with precise abundances for many individual elements. With uncertainties of $\sigma_{[X/{\rm Fe}]}$ and $\sigma_{\rm [Fe/H]} \simeq 0.05$ for 10 elements measured for $>10^4$ stars, the APOGEE DR12 spectra may be the first well-suited data set to put this idea into practice. We find that even APOGEE data offer only $\sim 500$ independent volume elements in the 10-dimensional abundance space, when we focus on the $\alpha$-enhanced Galactic disk. We develop and apply a new algorithm to search for chemically homogeneous sets of stars against a dominant background. By injecting star clusters into the APOGEE data set we show that chemically homogeneous clusters with masses $\gtrsim 3 \times 10^7 \, {\rm M}_\odot$ would be easily detectable and yet no such signal is seen in the data. By generalizing this approach, we put a first abundance-based constraint on the cluster mass function for the old disk stars in the Milky Way.
\end{abstract}


%
%
%
%
%
%

\section{Introduction}
\label{sec:introduction}

The Milky Way offers a unique opportunity to understand how disk galaxies form, in particular when and where, and in which types of aggregates, or clusters, they formed their stars.  As star cluster masses depend on gravitational instabilities \citep[e.g.,][]{esc08}, the cluster mass function (CMF) indirectly probes the dynamical state of the Milky Way disk over cosmic time. 

At least during the intensely star-forming, early phases of the Milky Way, the majority of stars are believed to form in clusters \citep{krui12,ada15}. Most of these are rapidly disrupted and dispersed throughout the Galaxy \citep[e.g.,][]{ode03,kop10,dal15}, for a brief while appearing as moving groups \citep[e.g.,][]{des07a,des13,bub10}. Once the phase space information as a common birth marker is lost, chemical tagging, first proposed by \citet{fre02}, may still betray the common birth origin of stars through their exceptional similarity in element abundances. Stars originating from the same star cluster are believed to be homogeneous in their chemistry \citep[e.g.,][]{des07b,kop08,des09,tin12b,fen14,fri14}. Since the photospheric element abundances, at least for elements heavier than sodium, are invariant throughout their lifetime, they are permanent tags of the stellar birth origins. 

A broad goal of chemical tagging is to reconstruct the stellar CMF, i.e. the relative distribution of (chemically homogeneous) stellar cluster masses when they formed. Although we can investigate the present-day CMF through young star clusters and massive surviving clusters \citep[e.g.,][]{bic03,lad03,por03,bra08,bor11,bra12}, the Milky Way's CMF in the past is unknown. The key idea in chemical tagging is that massive chemically homogeneous clusters show up as discernible clumps in the multi-dimensional abundance space \citep[e.g.,][]{bla10a,tin15}. 

Besides understanding the CMF, chemical tagging is deemed an essential tool to understand the effect of radial migration in Galactic evolution \citep[e.g.,][]{sch09,gou15}. \citet{sel02} first proposed that stars could migrate significantly from their radial position when resonate with spiral/bar structures \citep[see observational evidence from][]{hayw08,loe11,kor15}. Although simulations concur to the analytic calculations \citep{ros08,ros12,bla10b,min10,dim13,hal15}, direct observational evidence of radial migration remains controversial. The ability to recover dispersed star clusters would be fundamental in quantitatively constraining radial migration models.

In recent years, the idea of chemical tagging has garnered more attention.  Large spectroscopic surveys, including RAVE \citep{ste06}, APOGEE \citep{zas13}, GALAH \citep{des15} and Gaia-ESO \citep{ran13}, are being carried out. Results from these surveys have demonstrated the power of using elemental abundance patterns to identify distinct stellar populations in the Milky Way. \citet{mar15} found young $\alpha$-enhanced stars that are difficult to explain within current models of the evolution of the Milky Way. \citet{mas15} and \citet{hay15} showed that the C/N ratio and the $\alpha$-elements are good indicators to separate the thin and thick components of the Galactic disk. Schiavon et al. (submitted) found bulge stars that show abundance patterns similar to globular clusters. These studies focus on finding populations of stars via their abundance patterns, which is a ``weak'' form of chemical tagging. The goal of the ``strong'' form of chemical tagging is to identify stars that were born from the same molecular cloud. When we refer to chemical tagging in this paper, we only refer to this ``strong'' form. In this context, due to these exciting new opportunities, many preparatory works started to explore the capabilities of these surveys and the idea of chemical tagging. For example, \citet{tab12,tab14,mit13,mit14,blan15,mac15} proposed various schemes and performed numerical experiments on separating open clusters/moving groups in abundance space.

\citet{tin15} explored the feasibility of chemical tagging by exploring a grid of Galactic evolution parameters. They found that identifying individual clusters through chemical tagging is generally challenging with the on-going surveys. Clumps that show overdensities in abundance space are usually made up of many smaller clusters, i.e., the background contaminants are non-negligible in clump search. One might be able to associate a detected clump as a single disrupted cluster only when the background density in abundance space is low. They showed that a low background can be achieved by either studying a subpopulation that occupies a large volume in abundance space, small abundance uncertainties or a large number of independent elements. Despite all these complications to reconstruct individual clusters, they argued that we can still statistically reconstruct the CMF through the clumpiness in abundance space. The main goal of this paper is to apply that idea to the APOGEE data \citep{hol15}. 

In \S\ref{sec:sample}, we characterize the APOGEE sample that we explore in this study. In \S\ref{sec:method}, we introduce our clump search method. The key is to define a robust search sphere that has the highest signal-to-background ratio possible. In \S\ref{sec:application}, we apply this method to the APOGEE DR12 data. We discuss how this result can be applied to obtain a tentative constraint on CMF. We conclude in \S\ref{sec:conclusion}. By comparing the signal-to-background contrast observed in the data to the simulated contrast from injected clusters, we will argue that no chemically homogeneous clusters more massive than $3 \times 10^7\,{\rm M}_\odot$ have formed in the $\alpha$-enhanced disk.

%
%
%
%
%
%

\section{APOGEE sample properties}
\label{sec:sample}

\begin{figure}
\includegraphics[width=0.45\textwidth]{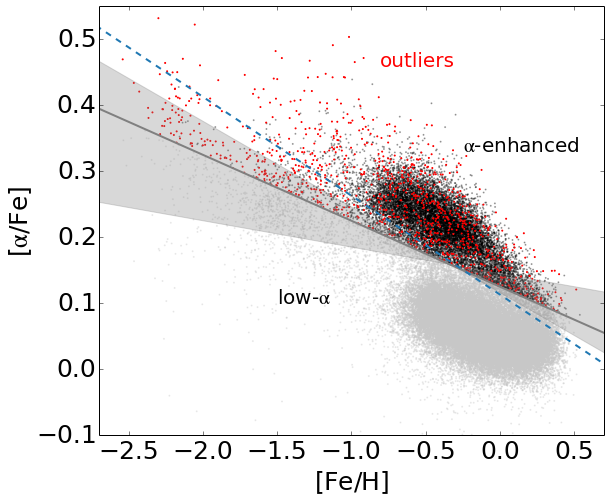}
\caption{Black symbols show 13,000 APOGEE stars selected for this study. We only consider $\alpha$-enhanced stars, as the volume they occupy in 10-dimensional abundance space is larger than the low-$\alpha$ sequence, so the background will be lower, and therefore detection is more likely. The red symbols show the $\sim 7\%$ of outliers in the 10-dimensional abundance space that we do not include in the sample. After we discard these outliers, the 10-dimensional empirical distribution in abundance space is better modeled by an ellipsoid and is easier to deconvolve (see \S\ref{sec:deconvolve}). The solid black line shows the fiducial selection cut. We also examine that the results in this paper are not sensitive to our data selection. If we choose a selection cut within the shaded region, the results in this paper still hold. The results only change more dramatically if we choose a much steeper cut such as the blue dashed line (see text for details).}
\label{fig:alpha-selection}
\end{figure}

We adopt the APOGEE DR12 publicly available sample \citep{hol15}. Similar to \citet{hay15}, we consider stars with all elements measured and with reliable abundances, i.e., 4,000 $< T_{\rm eff} <$ 5,500, $1 < \log g < 3.8$ and signal to noise ratio $> 80$. Stars from APOGEE that satisfy these criteria are plotted in Fig.~\ref{fig:alpha-selection}. In this study, we only focus on the $\alpha$-enhanced disk as defined via the cut shown in Fig.~\ref{fig:alpha-selection}. We focus on this subsample as their chemical/spatial modeling is likely to be more straightforward \citep[see \S\ref{sec:conversion} and also see][]{bla15,tin15}. Furthermore, the Milky Way was likely kinematically hotter and more turbulent in the first few billion years \citep[e.g.,][]{bou09,kru12,bir13}. High redshift ($z \sim 2$) extragalactic studies have revealed the existence of massive star-forming clumps in star-forming galaxies \citep[e.g.,][]{liv12,gen13}. As a result, star clusters within the $\alpha$-enhanced disk could be more massive and if so would provide a strong signal in abundance space. The $\alpha$-enhanced disk also occupies a larger volume in abundance space, i.e., lower background density of stars, which guarantees clumps a better contrast to the background in abundance space. 

We checked that our main result presented in this study, namely  there is no cluster more massive than $3\times 10^7 \, {\rm M}_\odot$ formed in the Milky Way, still holds at least to the $1\sigma$ level (see \S\ref{sec:CMF}), if we choose a selection cut within the shaded region as shown in Fig.~\ref{fig:alpha-selection}. The result only changes more dramatically if we choose a much steeper cut such as the blue dashed line. In this case, we discard too many low density regions where most chemical tagging signals reside (see \S\ref{sec:kernel} and \S\ref{sec:chemtag}), and we can only rule out clusters $\gtrsim 10^8 \, {\rm M}_\odot$. Using the fiducial selection cut, in total, the $\alpha$-enhanced sample has 14,002 stars, as shown in the red and black symbols in Fig.~\ref{fig:alpha-selection}. For reasons and selection criteria that will become clear in \S\ref{sec:deconvolve}, we further discard the 7\% of most outlying stars, shown as red symbols in Fig.~\ref{fig:alpha-selection}, and end up with a final sample of 13,000 stars.

Without further kinematic modeling, it is hard to disentangle the halo stars from the disk stars. But the elimination of outliers as shown in Fig.~\ref{fig:alpha-selection} culls most of the metal-poor stars with [Fe/H]$\,<-1$ and therefore, the bulk population in this study should not be contaminated much by the halo stars. We also performed the same analysis only considering stars with [Fe/H]$\,>-1$ and found that the results remain qualitatively the same. For the potential bulge contamination, we find that among the $13,000$ stars, only $3\%$ of them satisfy the bulge stars criteria with $l < 22$, $|b| < 15$ and $R_{\rm gc} < 3 \, {\rm kpc}$, where $R_{\rm gc}$ is the isochrone Galactic radius derived in \citet{hay15}. Therefore, we will assume throughout this study that the sample only consists of disk stars.

To perform the chemical tagging experiment, we want to consider as many elements as possible in order to maximize the volume in abundance space. In this case the background becomes more diluted, and the signals will therefore have a better chance of standing out from the background. In total, APOGEE measures 15 elements. However, as discussed in \citet{hol15}, Na, Ti, V abundances might not be reliable yet in the current release. Furthermore, C and N are expected and seen to evolve through stellar evolution due to the post main sequence dredge up \citep[e.g.][]{kar14,mas15,ven15}. Therefore, it is complicated to relate C and N to their primitive abundances when the stars formed. Discarding these 5 elements, we consider 10 elements in this study, namely Al, K, O, S, Mg, Si, Ca, Mn, Ni, Fe. Next, we want to define an abundance space out of these 10 elements. As discussed in \citet{tin12a}, chemical tagging is better performed in [$X$/Fe] instead of [$X$/H]. [$X$/H] strongly correlate with each other. It is harder to observe the subtle variants among clusters in an abundance space spanned by [$X$/H]. Therefore, in this study, we consider an abundance space spanned by [Fe/H] and 9 [$X$/Fe] from elements beside iron. We denote a vector in this 10-dimensional space to be {\bf X} in this study.

As we will discuss in more detail in \S\ref{sec:method}, to find a chemically homogeneous cluster in abundance space we first need to understand the typical volume that such a cluster occupies, after accounting for the measurement uncertainties that will dominate over the intrinsic scatter. Therefore, to estimate the typical volume, we will evaluate differential uncertainties, or measurement precision (not accuracy), $\sigma_{\bf X}$ and their correlations, i.e., the empirical point spread function of a chemically homogeneous cluster in abundance space. We estimate this co-variance matrix from known clusters in the APOGEE data and refer to the resulting matrix as the ``cluster kernel'' below.

We consider the DR10 clusters classification \citep{mes13} since the DR12 classification is yet to be released. We cross-match the DR10 cluster member identities with DR12 and adopt the element abundances from the DR12 release. We only consider clusters with more than 10 cluster members. Three open clusters (NGC6819, NGC2158, M67) satisfy this criterion. Noting that all these clusters are metal-rich with [Fe/H] $\gtrsim -0.1$ and the possibility that abundance determination could be worse at lower metallicity, we also adopt one of the more metal-rich globular clusters, M107  (${\rm [Fe/H]} \simeq -1$). We fit M107 members with two Gaussians distributions in the 10-dimensional abundance space to eliminate any possible secondary population from this globular cluster. Including M107, we have a total of 77 cluster members. The primary population of M107 shows measurement uncertainties consistent with other open clusters. Restricting ourselves to the three metal-rich open clusters results in a slightly smaller $\sigma_{\bf X}$. Therefore, including M107 only makes our results more conservative (see \S\ref{sec:detection-2D}). We subtract the element abundances of each cluster by their means to center clusters at the zero origin. The co-variance matrix of these 77 stars is estimated. This matrix defines an ellipsoid that a typical chemically homogeneous cluster occupies. For each element, we find that $\sigma_{\bf X} \sim 0.05-0.06 \, {\rm dex}$; this multivariate Gaussian sets the effective volume that homogeneous clusters occupy in the observed abundance space. This estimate is consistent with \citet{hol15} (see table 6 in the paper, but note that they show measurement uncertainties in [$X$/H], but here we evaluate uncertainties in [$X$/Fe] + [Fe/H]). Due to the small sample size of cluster members, we bootstrap the cluster sample and find that the uncertainty on this $\sigma_{\bf X}$ estimate is about $20\%$. A larger sample of cluster calibrators would be very helpful as a precise estimate of the cluster kernel is a key ingredient in any chemical tagging analysis.

%
%
%
%
%
%

\section{Method}
\label{sec:method}

In this section, we will describe the challenges in abundance clump searches and our clump search method. Although various schemes have been proposed to separate open clusters/moving groups \citep[e.g.,][]{sha09,blan15,mac15} in abundance space, a question often not discussed is the estimation and inclusion of background contaminants. Simulations from \citet{tin15} showed that the background contaminants can be a critical limiting factor in chemical tagging experiments. After extensive experimentation we found that most proposed techniques, such as K-means, Gaussian mixture models, and minimal spanning tree, are only effective in separating clumps in the limit of a small background or when the background can be easily estimated.

Due to this limitation, we have developed a simple new method\footnote{Our method can be regarded as a variation of density-based nonparametric clustering techniques.} geared toward regimes where the background is dominant and has a complex topology in abundance space \citep[read also][]{eve10}. The key to our method consists of two parts that we will explain in \S\ref{sec:kernel} and \S\ref{sec:background}. First, we need to estimate the local density. As we will discuss in more detail in \S\ref{sec:background}, we define the local density to be the number of stars within a search sphere. The search sphere that we use to distinguish signals from the background should be sufficiently large. It should include a large fraction of a chemically homogeneous cluster but avoid being too wide and should not include too many background contaminants. Secondly, the abundance space distribution is not uniformly distributed. To estimate the detection significance, we have to estimate the expected background at each location.

\begin{figure}
\includegraphics[width=0.45\textwidth]{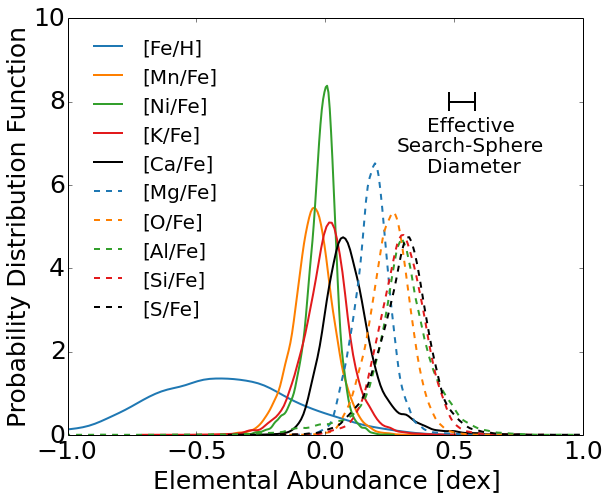}
\caption{Distributions of all 10 elements in this study. By comparing these distributions to the effective search sphere diameter (see \S\ref{sec:kernel} for details), we find the abundance distributions in each dimension to have standard deviation widths that are typically $\sim 1-3$ times of the effective search sphere radius. The relatively small volume in abundance space compared to the search sphere volume highlights the main challenge in chemical tagging.}
\label{fig:kernel-curse}
\end{figure}

%
%
%
%
%
%

\subsection{Abundance space search sphere}
\label{sec:kernel}

In \S\ref{sec:sample}, we derived the empirical multivariate Gaussian distribution that a typical cluster occupies in abundance space. This distribution defines an ellipsoidal distribution in the 10-dimensional abundance space. Since an optimal search sphere should include a high ratio of cluster objects compared to the background contaminants, the optimal search volume, or search sphere, should follow the same uncertainty ellipsoid. However, it is not convenient to operate with a tilted ellipsoid because a simple Euclidean distance from the center alone is not sufficient to determine whether a star is included in the ellipsoid. Therefore, we linearly transform the abundance space such that the cluster kernel becomes a unit Gaussian distribution. We emphasize that the transformation is only to make calculations more straightforward, leaving all astrophysical implications invariant.

We can now determine an appropriate radius for the search sphere in these coordinates. A unit radius is not a good choice for the search sphere even though clusters follow a unit Gaussian distribution in the transformed coordinates: a box with 2 dex in width in each dimension only captures $(68\%)^{10} = 2\%$ of the clump. Since a unit n-sphere is strictly included in this box, one can show that a unit n-sphere encapsulates an even smaller fraction, $0.02\%$, of the data. This curse of dimensionality\footnote{Techniques to compactify of dimensions, such as PCA \citep[e.g.,][]{tin12a}, do not mitigate this problem because the density of background contaminants also increases accordingly in the compactified space.} implies that to capture, for example, 68\% or $1\sigma$ of the cluster members, we require a search sphere with a radius larger than $1 \, {\rm dex}$ in the transformed abundance space. In fact, mathematically, the inclusion fraction of a unit Gaussian within an n-sphere follows the $\chi^2$-distribution. The analytic formula of a $\chi^2$-distribution shows that an n-sphere of $3.4 \, {\rm dex}$ in radius is needed.

One important parameter that will determine the difficulty of chemical tagging detections is the number of separate ``chemical cells'' in the abundance space \citep{fre02,tin15}. The number of chemical cells is the ratio between the abundance space volume spanned by the data and the typical volume of a search sphere. We will evaluate this number later. But for now, one way to visualize the number of chemical cells is to compare the effective diameter of the search sphere to the spread in each element abundance. We illustrate this comparison in Fig.~\ref{fig:kernel-curse}. We find that an n-sphere with a radius of $3.4 \, {\rm dex}$ in the transformed abundance space corresponds to an ellipsoid with an effective radius of $0.05$ dex in the original abundance space. The effective radius is defined such that an n-sphere with this radius contains the same volume as the ellipsoid. Due to the large dimensionality, we find that if we do not take into account the covariances of the cluster kernel, i.e., if we were to use an n-sphere in the original abundance space, instead of a tilted ellipsoid defined from the cluster members, we estimate that the search volume would be 10,000 times larger and the search sphere would include too many background contaminants.

Even with this optimized search ellipsoid in the original abundance space, as shown in Fig.~\ref{fig:kernel-curse}, the distribution of each element is typically only 1-3 times the effective search sphere diameter. As a result, it is not possible to search for clusters in the core region of the chemical distribution. In this region, the search sphere would include too many background contaminants. The chemical tagging signals are most likely to come from the peripheral regions of the chemical distribution where the background contaminants are not dominant \citep[also see][]{bla10a,kar12,bla15}. Nonetheless, in a 10-dimensional space, the ``surface-to-core'' ratio is very large. Therefore, there is a reasonable chance of finding clumps in the peripheral regions. As we will show in \S4, all chemical tagging signals indeed come from the peripheral regions.

\begin{figure*}
\includegraphics[width=1.0\textwidth]{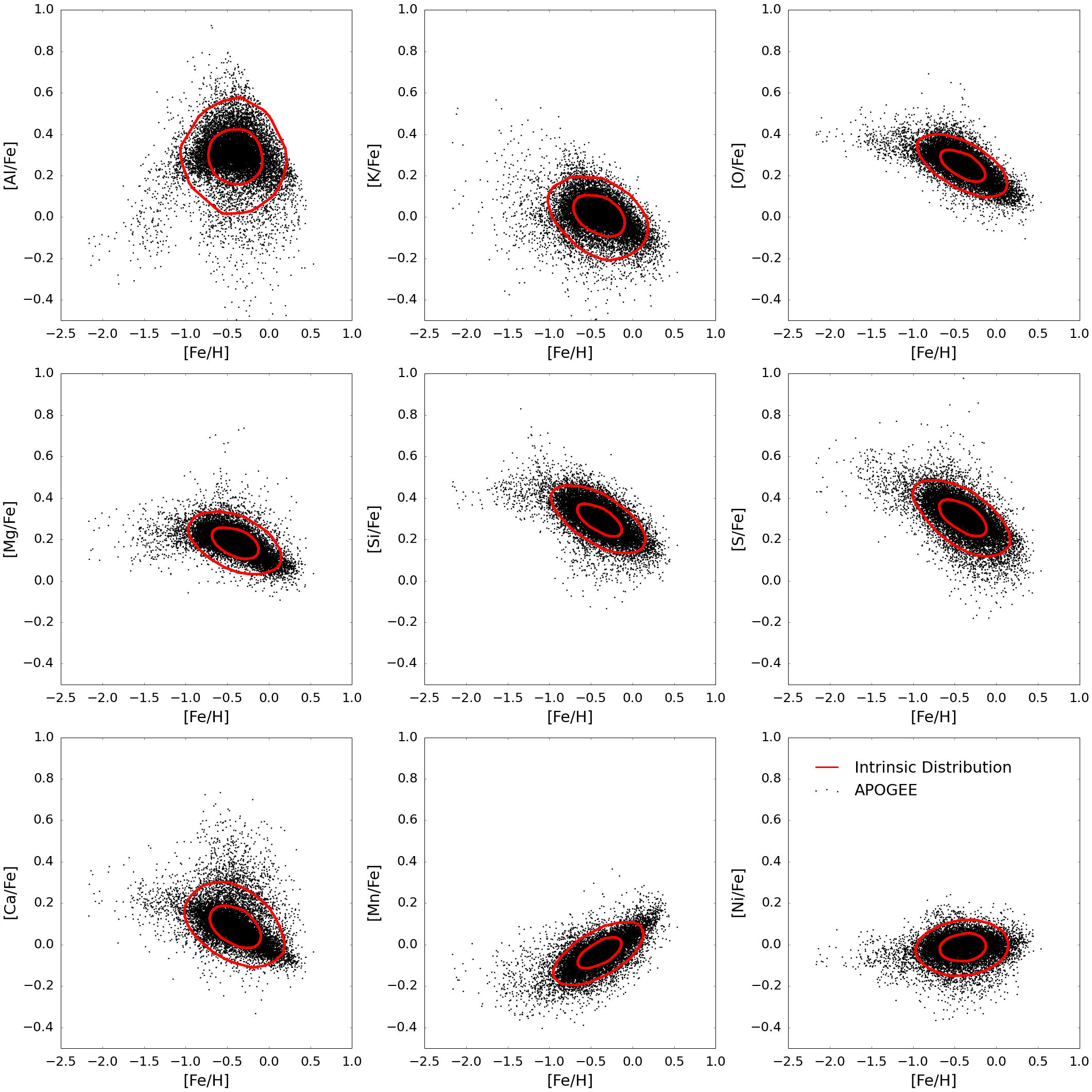}
\caption{Deconvolved (intrinsic) model of the 10 elements in this study, projected onto [$X$/Fe]-[Fe/H] planes. The black symbols show the APOGEE sample adopted, illustrating the empirical distribution. The red contours show the 50 and 90 percentiles of the projected intrinsic distributions. When injecting mock clusters, we draw their center locations from the intrinsic distribution model instead of the empirical APOGEE distribution. Note that although an ellipsoid model provides an acceptable fit to the data, in some cases the abundance distribution is influenced by non-ellipsoidal structures.}
\label{fig:deconvolved-distribution}
\end{figure*}

\vspace{1cm}

%
%
%
%
%
%

\subsection{The intrinsic abundance distribution of the $\alpha$-enhanced sample}
\label{sec:deconvolve}

We now proceed to deconvolve the observed abundance distribution.  The intrinsic abundance distribution is required in order to inject mock clusters into the observed APOGEE dataset.

In the previous section, we transformed abundance space such that clusters follow a 10-dimensional unit Gaussian distribution. However, in such coordinates, the overall chemical distribution of the $\alpha$-enhanced disk will still presumably show co-variances among the different coordinate directions because the transformation is only to normalize the cluster point spread function and makes no assumption on the global distribution. Deconvolving such a co-variant distribution directly in 10-dimensions is computationally prohibitive. Therefore, we further rotate the transformed abundance coordinate system to eliminate the co-variances such that the joint 10-dimensional abundance distribution of the $\alpha$-enhanced disk can be approximated by a product of 10 marginal distributions. Since the cluster kernel, reflecting the measurement uncertainties, is already isotropic in the transformed abundance space, it remains unaffected by further rotation. Upon this rotation, the deconvolution task simplifies to ten independent, one-dimensional deconvolutions on the marginal distributions.

However, this approach only works if the actual ensemble abundance distribution can be well approximated by its marginal distributions after rotation, i.e. if it does not show significant curvature in abundance space. If we consider all 14,002 $\alpha$-enhanced stars in the sample, we find that this is not a good approximation: the 10-dimensional abundance space has a less regular topology, dominated by a small fraction of outliers in abundance space, as shown in red symbols in Fig.~\ref{fig:alpha-selection}. To look for these outliers, we perform a 10-dimensional kernel density estimation, using a unit Gaussian distribution as the smoothing kernel. We rank the data points according to their local density in the kernel density estimation and discard the most outlying 1,002 stars. We check that upon discarding these outliers (7\%), injecting clusters according to the joint distribution gives similar statistics to injecting clusters according to product of marginal distributions. We emphasize that discarding outliers shrinks the peripheral regions and makes detection more unlikely. The main purpose of this study is to put an upper limit on the maximum cluster mass (see \S\ref{sec:CMF}), discarding outliers only makes our estimate more conservative. We also check that these outliers are not particularly clumped in abundance space and hence are unlikely to be chemical tagging detections.

After breaking down the joint distribution to its marginal distributions, $P_{{\rm convolved},i}$, we model each marginal distribution with the sum of two Gaussians \citep[see also][]{mcl00,eve10},
\begin{eqnarray}
\label{eq:composite-gaussian}
P_{{\rm convolved},i} (x_i|\mu_1,\mu_2, \sigma_1^2,\sigma_2^2,f) \qquad\qquad\qquad \nonumber \\
\sim (1-f) \mathcal{N}(x_i|\mu_1,\sigma_1^2) + f \mathcal{N}(x_i|\mu_2, \sigma_2^2).
\end{eqnarray}

\noindent
where $\mu$ and $\sigma$ are the means and standard deviations of the Gaussian distributions, $f$ shows the relative contribution from each distribution. We require a two-component Gaussian because the marginal distributions often show a core region and a broad wing region. Fitting a single Gaussian will underestimate the total area of low-density wings. As discussed in \S\ref{sec:kernel}, the wing regions are the most valuable parts of a chemical distribution as they have the highest chance to detect clumps. In the model, we allow centers of the two Gaussian distributions, $\mu_1$ and $\mu_2$, to be different to account for an asymmetric distribution. We found that this model provides an excellent fit to each 1D marginalized distribution (although the joint distribution fits are slightly less satisfactory as we will explain below) and the deconvolution can be done analytically. We model the intrinsic chemical distribution $P_{\rm intrinsic}$ to be
\begin{equation}
P_{\rm intrinsic} ({\bf X}) = \prod_{{\rm dim}=1}^{10} P_{{\rm intrinsic},i}(x_i),
\end{equation}

\noindent
where 
\begin{equation}
P_{{\rm intrinsic},i} (x_i) = P_{{\rm convolved},i} (x_i|\mu_1,\mu_2,\sigma_1^2-1^2,\sigma_2^2-1^2,f).
\end{equation}

Fig.~\ref{fig:deconvolved-distribution} shows the intrinsic chemical distribution derived according to the method above. We will use this model to draw mock data of hypothetical clusters, whose abundance probability is drawn from the ensemble distribution. We caution that the 10-dimensional ellipsoidal model does not give a perfect fit to the data, despite the fact that we have eliminated 7\% of the outliers. For instance the [Ca/Fe] vs. [Fe/H] distribution, as illustrated in Fig.~\ref{fig:deconvolved-distribution}, shows a more complex morphology than an ellipsoidal model. In the ideal case, we would draw hypothetical clusters from a deconvolved distribution that displays similar intricate morphology. However, deconvolving such intricate distribution is computationally intractable in 10-dimensional. Nonetheless, we checked that injecting clusters according to the convolved ellipsoidal model shows similar statistics to the empirical convolved distribution. Therefore, in this study, we make the assumption that injecting cluster according to the deconvolved ellipsoidal model gives similar statistics if we were to inject according to the real deconvolved distribution.

%
%
%
%
%
%

\subsection{Detection significance}
\label{sec:background}

So far, we have defined an operative search sphere with radius $r = 3.4 \,{\rm dex}$ to look for overdensities in the transformed abundance space. We know that the overall abundance distribution, i.e. the background, is not uniform \citep[e.g.,][]{edv93,bar05,red06,ben14}. The absolute number of stars within the search sphere is therefore not particularly informative. We want to find regions where the local density within a search volume is significantly higher than its vicinity regions. Therefore, to define the detection significance, we need two ingredients: a local density estimation at each location and the corresponding local background estimation at this location. 

\begin{figure}
\includegraphics[width=0.48\textwidth]{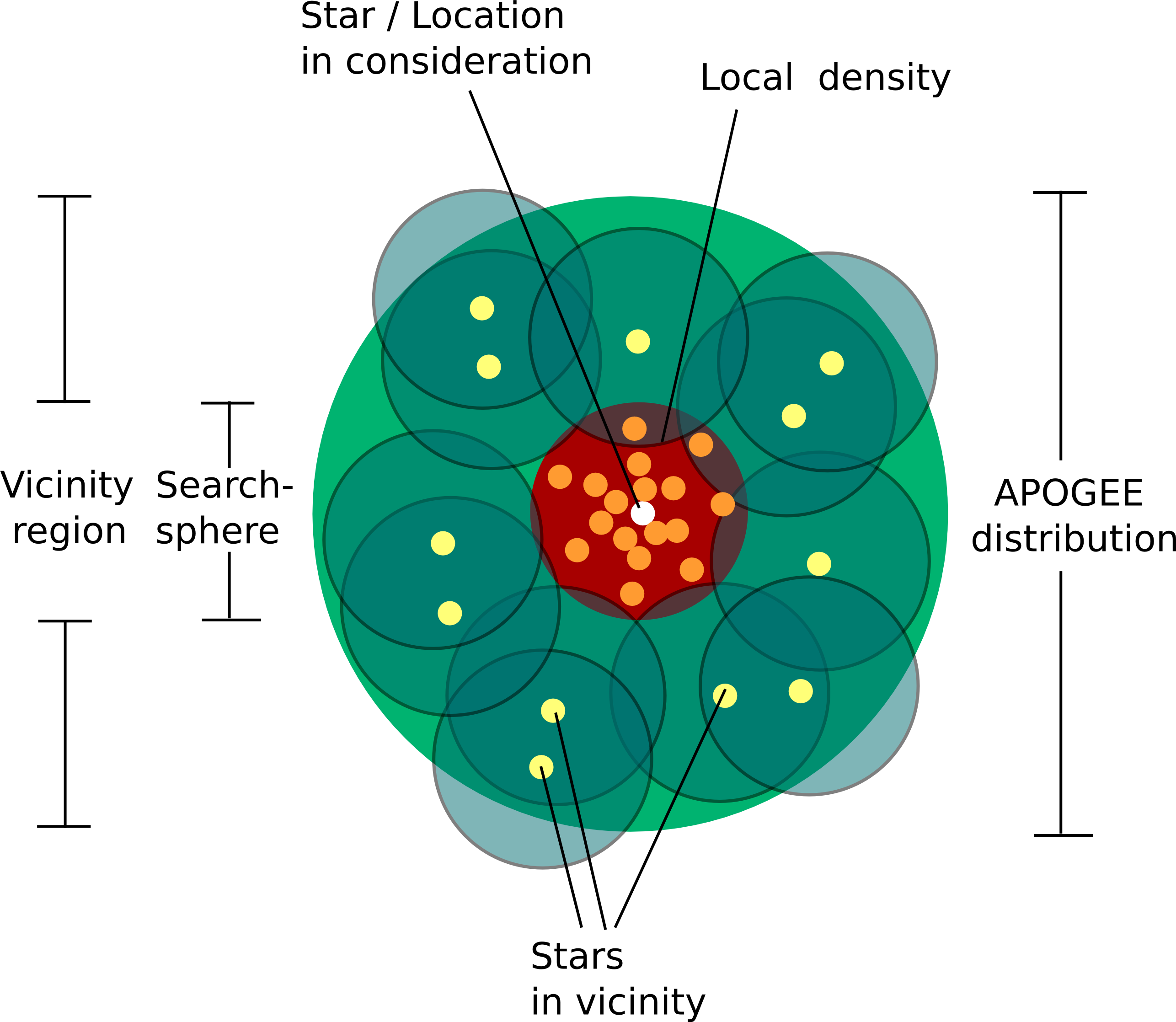}
\caption{Schematic illustration of our clump search method. At each star $s$ (white symbol), we evaluate the number of stars (orange symbols) within the search sphere (red shaded region) that we denote as $n_s$. The vicinity background, $\langle n_{s'} \rangle$ is calculated by averaging other $n_{s'}$, as shown in the blue shaded regions, where $s'$ are all stars (yellow symbols) that are in the vicinity region (green shaded region). The vicinity region is defined to be the region outside the search sphere but inside two times the search sphere. The detection significance is then defined as $\sigma_{\rm detect}(s)=(n_s - \langle n_{s'} \rangle)/\sqrt{\langle n_{s'} \rangle}$.}
\label{fig:clumps-method}
\end{figure}

Fig.~\ref{fig:clumps-method} shows a schematic illustration of our clump search method. For each star $s$, we define the local density of this star, $n_s$, to be the total number of stars located within $r$ distance from this star. Throughout this study, we only consider stars with $n_s > 10$ to avoid the large Poisson fluctuation at small $n_s$. We estimate the vicinity background, $\langle n_s \rangle$ to be the average of $n_{s'}$ where $s'$ are all stars located within a distance of $r - 2r$ from star $s$. We define the detection significance to be $\sigma_{\rm detect}(s) = (n_s - \langle n_s \rangle)/\sqrt{\langle n_s \rangle}$, which measures the deviation of the local density from the vicinity background, in units of the Poisson uncertainty of the background. 

Although this is a sensible definition of detection significance, there is a complication. If we have a uniform background, provided there is no signal, $\sigma_{\rm detect}$ should center around zero. Unfortunately, this is not the case for an uneven background, especially for high dimensions. At a fixed point in an uneven background, there are always more vicinity regions that have lower densities (toward the valley) than regions that have higher densities (toward the core). As a result, we have $n_s > \langle n_s \rangle$ in general. This disproportion gets more severe toward the core as the background gradient gets steeper. This disproportion causes $\sigma_{\rm dense}(n_s)$ to be an increasing function of $n_s$. To overcome this shortcoming and to have $\sigma_{\rm detect}$ centered around zero, we calibrate $\sigma_{\rm detect}$ by the median of $\sigma_{\rm detect} (n_s)$ at each $n_s$. We denote the calibrated detection significance to be $\overline{\sigma}_{\rm detect}$ and use it to be our operative measure of detection significance in the following and apply this method to the APOGEE data.

\begin{figure*}
\includegraphics[width=1.0\textwidth]{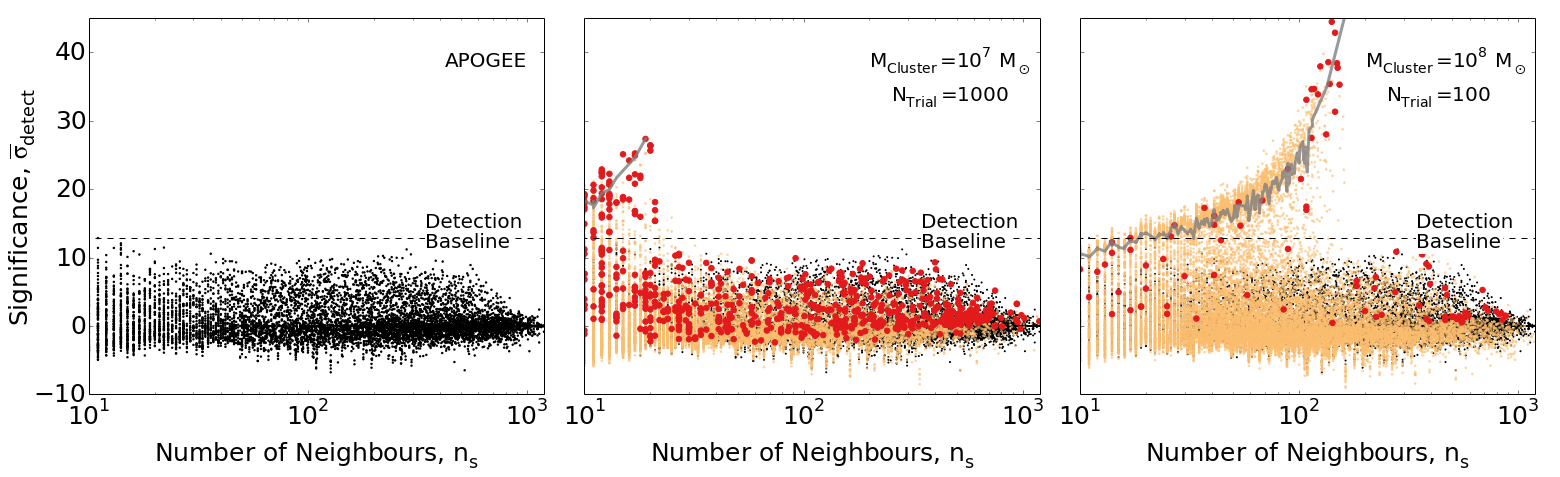}
\caption{Calibrated detection significance as a function of local density in abundance space. The left panel shows the sample observed values. The dashed lines show the observation detection baseline. The overlaid orange symbols in the middle and right panels show results from the injected $10^7 \, {\rm M}_\odot$ and $10^8 \, {\rm M}_\odot$ clusters, respectively. Since we calculate the local density centered at {\em each} star, there are numerous points for each cluster. We highlight the maximum deviation of each cluster with a bold red symbol. We compile results from 1000 and 100 trials for the middle and right panels. To demonstrate a typical detection significance distribution within a cluster, we link with solid gray lines all stars from the cluster containing the star with the highest detection significance. The middle and right panels show that if a $10^7 - 10^8 \, {\rm M}_\odot$ cluster formed in the past, some cluster members might show more deviations than the detection baseline.}
\label{fig:detection-criteria}
\end{figure*}

%
%
%
%
%
%

\section{Results}
\label{sec:application}

We now explore what we can learn about the number of chemical cells in the APOGEE survey, and about the presence of any clumps in abundance space, that may reflect chemically tagged remnants of disrupted clusters.

%
%
%
%
%
%

\subsection{The Number of chemical cells in the APOGEE observations of the $\alpha$-enhanced Galactic disk}
\label{sec:numberofcells}

The number of chemical cells is best estimated in the transformed (and rotated) coordinates where the global chemical distribution has no co-variances between different coordinates and a chemically homogeneous cluster can be represented by a unit Gaussian distribution. To calculate the number of chemical cells, let us estimate the global distribution to be a multivariate Gaussian with no co-variances and with standard deviations $\sigma_1, \sigma_2, \ldots, \sigma_{10}$ in the 10 transformed coordinates. The number of chemical cells, by definition, is the volume ratio between the global distribution over the cluster kernel. Since the cluster kernel has a unit width in all directions, the number of chemical cells in APOGEE can be estimated to be $(\sigma_1 \cdot \sigma_2 \cdots \sigma_{10})/(1 \, {\rm dex})^{10} \simeq 500$.

This estimate agrees with the prediction in \citet{tin12a} using principle components analysis. The lack of chemical cells despite having 10-dimensions is due the strong correlations among abundances, especially for the $\alpha$-capture elements and Fe peak elements. The small number of chemical cells emphasizes the challenges in performing chemical tagging with strongly correlated element abundances. Nonetheless, as we have discussed in \S\ref{sec:deconvolve}, the APOGEE abundance space has broad wings that are not captured in a single multivariate Gaussian. Therefore, in the analysis presented below, the signals drawn from a {\em composite} multivariate Gaussian, as described in equation~(\ref{eq:composite-gaussian}), are stronger than a simple multivariate Gaussian with 500 chemical cells.

%
%
%
%
%
%

\subsection{Relation between $N_{\rm inject}$ and $M_{\rm cluster}$}
\label{sec:conversion}

To understand whether a cluster will be detected or not, we first need to investigate the number of stars that we would sample in APOGEE from a cluster, given its zero age cluster mass $M_{\rm cluster}$. We denote the number of stars sampled to be $N_{\rm inject}$ and will use the one-to-one relation between $N_{\rm inject}$ and $M_{\rm cluster}$ in the following discussion. But this conversion is based on some critical assumptions that we will now explain. The relation between $N_{\rm inject}$ and $M_{\rm cluster}$ is one-to-one up to a Poisson uncertainty -- more massive clusters have more stars to begin with, and therefore will have more stars sampled in the survey.

In the limit where there is no radial migration, the relation between $N_{\rm inject}$ and $M_{\rm cluster}$ is simple and can be derived analytically. Assuming stars are azimuthally mixed in the annulus, the number of stars sampled from a cluster, $N_{\rm inject}$, can be approximated \citep[see][]{bla15,des15,tin15} to be
\begin{equation}
\label{eq:convert}
N_{\rm inject} = \frac{M_{\rm cluster}}{M_{\rm annulus}} N_{\rm APOGEE},
\end{equation}

\noindent
where $M_{\rm annulus}$ is the total stellar mass (including stellar mass loss) in the annulus and $N_{\rm APOGEE} =$ 13,000 is the APOGEE sample size in this study. This formula can be easily understood as the following. Assuming stars in the sample have an average stellar mass $\langle M \rangle \simeq 1 \, {\rm M}_\odot$, the ratio $N_{\rm APOGEE}/{M_{\rm annulus}}$ gives the stellar mass fraction within the annulus that we would sample in the survey. We denote this ratio to be the sampling rate. When multiplying the sampling rate by a cluster mass, the product gives the stellar mass, and thus the number of stars with $\langle M \rangle = 1 \, {\rm M}_\odot$, that we would sample from this cluster. 

However, radial migration modifies $M_{\rm annulus}$ in a complex way \citep[see details in][]{tin15}. Stars born outside the annulus could migrate into the annulus, and stars from the annulus could now appear to be outside the annulus. Due to this complication, to estimate $M_{\rm annulus}$, the APOGEE's selection function, as well as a robust Galactic chemical evolution model \citep[e.g.,][]{kob06,kob11,min13}, is needed. This is clearly beyond the scope of this paper.

To simplify the problem and to only derive a conservative limit on the CMF in \S\ref{sec:CMF}, we assume that the $\alpha$-enhanced disk is completely radially mixed. In other words, the current spatial location of a star is completely random and is independent of their birth radii. In this limit, a star in the sample can be any star from the $\alpha$-enhanced disk. Therefore, we have $M_{\rm annulus} = M_{\rm total}$, where $M_{\rm total}$ is the total stellar mass of the $\alpha$-enhanced disk. Although complete mixing is a crude assumption, it is likely to be reasonable for the $\alpha$-enhanced disk. For example, \citet{hay15} showed that there is a universal $\alpha$-trend irrespective of the Galactocentric radii. A natural explanation of this result is the stars in the $\alpha$-enhanced disk are well-mixed.  Moreover, the APOGEE sample covers a wide range of Galactocentric radius, with $R_{\rm gc} \simeq 3 \, {\rm kpc} - 15 \, {\rm kpc}$ \citep{bov14,nid14,hay15}. It should have sampled the $\alpha$-enhanced disk from a large fraction of the Milky Way. 

We emphasize that the complete mixing assumption gives a conservative limit on the CMF. In the case where the mixing is not complete, we would have sampled more stars from the same cluster, and hence, it would be easier to exceed the APOGEE baseline. On top of the complete mixing assumption, we also assume that the CMF is independent of element abundances, and hence the equation~(\ref{eq:convert}) applies universally to the whole abundance space. 

With these assumptions, we only need to properly estimate $M_{\rm total}$ and apply equation~(\ref{eq:convert}) to obtain a one-to-one relation between $N_{\rm inject}$ and $M_{\rm cluster}$. We assume that the $\alpha$-enhanced disk has an exponential scale length of $3 \, {\rm kpc}$ \citep[e.g.,][]{bov12} and consists of $10\%$ stellar mass observed in the solar neighborhood \citep[e.g.,][]{che12}. We adopt the stellar density in the solar neighborhood to be $38 \, {\rm M}_\odot {\rm pc}^{-2}$ \citep[e.g.,][]{fly06,bov13,zha13}, and the solar Galactocentric radius $R_0 = 8 \, {\rm kpc}$ \citep[e.g.,][]{ghe08,gil09,rei14}. These assumptions yield a present-day $\alpha$-enhanced disk stellar mass of $\sim 3 \times 10^9 \, {\rm M}_\odot$. Since the $\alpha$-enhanced disk is old, massive stars have long since evolved and died. To account for this, we consider a total stellar mass loss of $40 \%$ \citep[][assuming a Kroupa IMF]{con09}. Putting all these together, we have $M_{\rm total} \simeq 6 \times 10^9 \, {\rm M}_\odot$. We also derive the sampling rate of the current APOGEE sample to be $N_{\rm APOGEE}/M_{\rm total} = \frac{1}{5\times 10^5}$. On average, we would collect one star from a $5 \times 10^{5} \, {\rm M}_\odot$ cluster. We will defer the discussion on what this low sampling rate implies in \S\ref{sec:CMF}.

%
%
%
%
%
%

\subsection{Chemical tagging in APOGEE}
\label{sec:chemtag}

We apply the clump search method described in \S\ref{sec:method} to the APOGEE sample. The left panel of Fig.~\ref{fig:detection-criteria} shows $\overline{\sigma}_{\rm detect}$ as a function of $n_s$ of all 13,000 stars.  At face value, it is tantalizing to observe deviations $> 5 \sigma$. But we emphasize that the detection significance depends on the various assumptions made, such as the choice of search sphere radius, the minimum number of neighbors requirement, the detection significance calibration and the definition of vicinity region at each data point. Without further information such as stellar ages, it is difficult to confirm the origin of these clumps. Furthermore, as shown in \citet{tin15}, most clumps are comprised of many clusters sharing similar element abundances.

Due to these uncertainties, instead of interpreting these clumps as detections, we proceed by assuming the APOGEE dataset (left panel of Fig.~\ref{fig:detection-criteria}) to be the detection baseline. We inject mock clusters of different sizes into the data and estimate the detection significance of these injected objects. By forward modeling, we {\em rule out cases that are not consistent with the observation baseline}. The middle and right panels show $\overline{\sigma}_{\rm detect}$ of the injected objects. In the right panel, we combine results of 100 trials, where in each case we inject a $10^8 {\rm M}_\odot$ ($N_{\rm inject} \simeq 250$ stars) clump into the data. In the middle panel, we show the results of 1000 trials with $10^7 {\rm M}_\odot$ clusters ($N_{\rm inject} \simeq 25$ stars) injected. These two panels show that if $10^7-10^8 {\rm M}_\odot$ clusters have formed in the past, there is a reasonable chance that we would have detected larger deviations than the value observed. A cluster with $10^7 {\rm M}_\odot$ lies above the detection boundary about $\sim 7 \%$ of the time, and a cluster with $10^8 {\rm M}_\odot$ is detected about $\sim 30 \%$ of the time.

Not all high mass clusters will exceed the detection baseline. As shown in the middle and right panels, most clusters, especially at high $n_s$, blend into the background. To make robust statements, we now proceed to quantify the probability of a cluster exceeding the observation baseline. 

\begin{figure}
\includegraphics[width=0.5\textwidth]{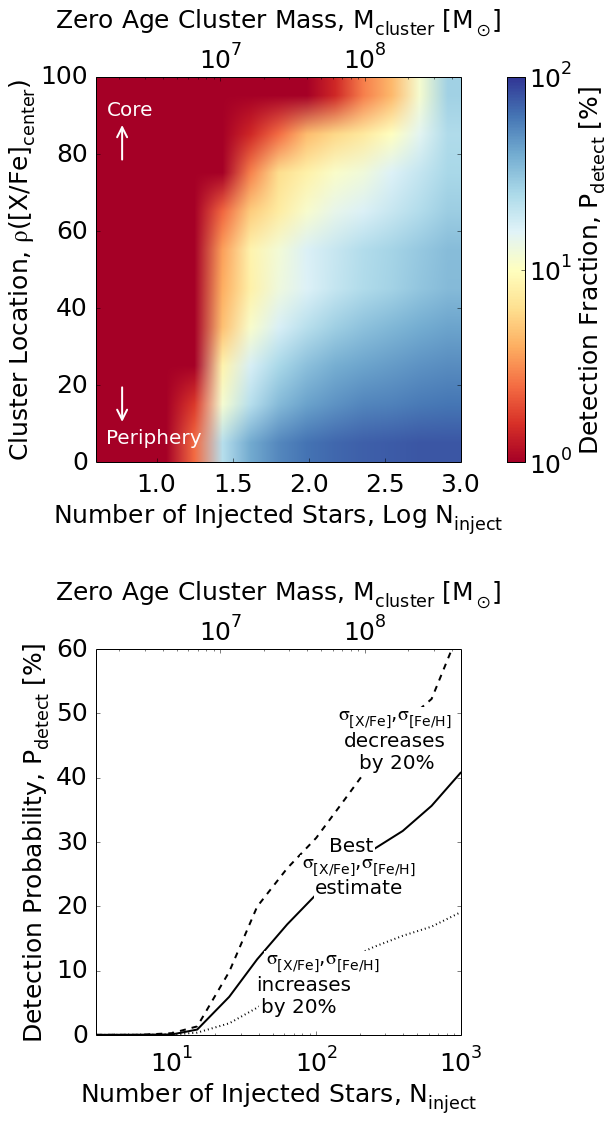}
\caption{Probability of an injected cluster showing more significant detection than the APOGEE data. This probability depends on two parameters: the number of injected stars per cluster and the cluster location in abundance space. The top panel shows the probability as a function of these two parameters. As the number of injected stars increases or the cluster is injected in a more peripheral region of abundance space, the chance to exceed the observation baseline increases. The bottom panel shows the probability marginalized over the cluster location, i.e., the probability of detecting a cluster of a certain cluster mass if the cluster location is randomly drawn from the intrinsic abundance distribution. The solid line shows result assuming the best cluster kernel estimation, $\sigma_{\bf X}$, as also applied to the top panel. The dashed and dotted lines show the marginal probability assuming $\pm 20\%$ statistical uncertainties of the $\sigma_{\bf X}$ estimate due to the small sample of cluster members (see \S\ref{sec:sample} for details).}
\label{fig:detection-proba}
\end{figure}

%
%
%
%
%
%

\subsubsection{Detection probability of individual clusters}
\label{sec:detection-2D}

In this section, we will quantify the probability of an injected clump exceeding the observation baseline.  There are two key parameters that determine this probability: (a) the number of stars injected as a clump, $N_{\rm inject}$, and (b) the cluster location in abundance space. As for the latter, the clump centers are drawn from the intrinsic distribution model described in \S\ref{sec:deconvolve}.  As for the former, we consider a grid of $N_{\rm inject}$, ranging from $3-1000$ stars with a step-size of $0.2$ in log scale. When injecting a mock cluster with $N_{\rm inject}$ stars, we allow a Poisson fluctuation of $\sqrt{N_{\rm inject}}$. The $N_{\rm inject}$ range in this study roughly corresponds to cluster masses of $10^6 {\rm M}_\odot - 5 \times 10^8 {\rm M}_\odot$. We run $10^4$ trials for each $N_{\rm inject}$ and find that the Monte Carlo uncertainty is negligible with this many trials. 

We model the cluster location by ranking all $10^4$ trials by their $P_{\rm intrinsic}({\bf X}_{\rm center})$ value. ${\bf X}_{\rm center}$ is the clump center location in abundance space. We put the ranking into a linear scale, which we will denote as $\rho({\bf X}_{\rm center}) \in [0,100]$, where

\begin{equation}
\rho({\bf X}_{\rm center}) \equiv \frac{\# \big({\rm trials} < P_{\rm intrinsic}({\bf X}_{\rm center}) \big)}{\#({\rm trials})}.
\end{equation} 

\noindent
If the cluster is located near the background dominated core, it has a higher value in $P_{\rm intrinsic}$ because the background density is very large, and we assign a high $\rho({\bf X}_{\rm center})$. Whereas, if the cluster is located at the peripheral regions, it has a lower $P_{\rm intrinsic}$ value since the background density is low, and we assign a low $\rho({\bf X}_{\rm center})$.

For each trial, we inject a clump and estimate the local density and vicinity background for all objects from the injected clump the same way in \S\ref{sec:background}. We define a clump to have exceeded the observational baseline if the maximum detection significance of this clump exceeds the baseline as demarcated by the dashed lines in Fig.~\ref{fig:detection-criteria}. We take the maximum significance of the whole clump because not all injected objects will have high deviations, as shown in the solid gray lines in Fig.~\ref{fig:detection-criteria}. Objects located near the surface of a clump will blend into the background. Only the objects near the clump center will have high deviations because the search sphere includes a large fraction of the clump.

The top panel of Fig.~\ref{fig:detection-proba} shows the probability of exceeding the APOGEE baseline. Among all trials that have a similar cluster location and a similar number of stars, we evaluate the fraction of them exceeding the baseline. The x-axis shows $N_{\rm inject}$ and the y-axis shows the cluster location quantified by $\rho({\bf X}_{\rm center})$. This panel illustrates that as the number of injected objects increases or the cluster location is increasingly toward the peripheral regions, the chance of exceeding the baseline improves, consistent with our intuition from Fig.~\ref{fig:detection-criteria}.

The solid line in the bottom panel of Fig.~\ref{fig:detection-proba} shows the probability marginalized over the cluster location, i.e., the probability of a cluster exceeding the APOGEE baseline as a function of its cluster mass if the cluster location is randomly drawn from the intrinsic abundance distribution. The marginalized probability shows that clusters less massive than $10^7 \, {\rm M}_\odot$ have negligible chances of exceeding the baseline, but clusters more massive than $\sim 10^7 \, {\rm M}_\odot$ begin to show tension with the deviations observed in APOGEE. The bottom panel also illustrates that even for a cluster as massive as $\sim 5 \times 10^8 \, {\rm M}_\odot$ ($N_{\rm inject} \simeq 1000$), only about half of the time will a cluster exceed the baseline. The lack of significant detection from the other half is not unexpected. As illustrated in the right panel of Fig.~\ref{fig:detection-criteria}, if a cluster is located in the core region (i.e., high $n_s$), the background becomes dominant. In this regime, most objects within the search sphere come from background contaminants. Therefore, in the core region, the signal tends to be overwhelmed by the background, regardless of the cluster size. 

Recall that our estimate of the cluster kernel $\sigma_{\bf X}$ has an uncertainty of $20\%$ due to the small number of cluster stars. Therefore, the concentration of our injected clusters could be off by the same amount. We also explore how this uncertainty might change our results. The dashed and dotted lines show results in cases where our cluster concentration estimate is off by $20\%$. The dashed line shows the result assuming chemically homogeneous clusters are intrinsically tighter in abundance space by $20\%$. With a more concentrated signal, the signal will have a better contrast over the background. Therefore, clumps are easier to detect and the probability in Fig.~\ref{fig:detection-proba} increases. However, if clusters are more widely spread, they are more likely to blend into the background. Therefore, the chance of detection decreases, as shown in the dotted line. We defer more detail discussions on how this uncertainty changes our conclusion to \S\ref{sec:CMF}.

\begin{figure*}
\includegraphics[width=1.0\textwidth]{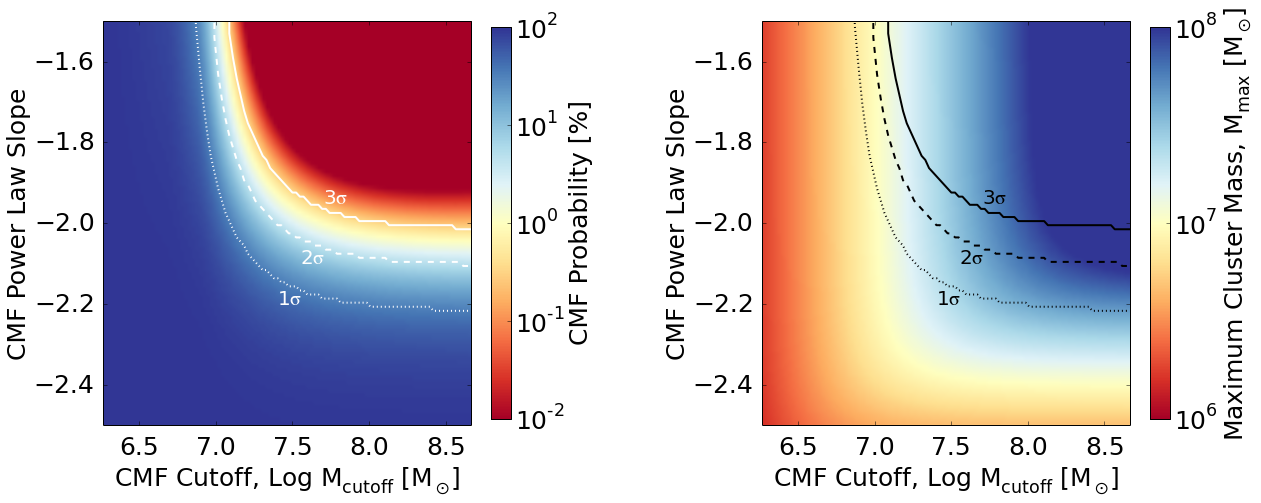}
\caption{Constraint on the $\alpha$-enhanced disk CMF. The left panel shows the probability of a CMF being consistent with the APOGEE DR12 data as a function of the two CMF parameters, the power-law slope and the upper mass cutoff. The dotted, dashed, and solid lines show the $1-3$ sigma limits, respectively. Unless the CMF power-law slope is steeper than $-2$, a cluster mass cutoff $\gtrsim 3 \times 10^7 \, {\rm M}_\odot$ is largely ruled out. The right panel shows the maximum cluster mass, $M_{\rm max}$, for different CMFs, such that the expected number of clusters $N_{\rm cluster}(> M_{\rm max}) > 1$ (see \S\ref{sec:CMF} for details). We overplot the $1-3$ sigma limits calculated from the left panel. The right panel shows that, in most cases, there is on average less than one cluster with $\gtrsim 3 \times 10^7$ formed in the Milky Way.}
\label{fig:CMF-limit}
\end{figure*}

%
%
%
%
%
%

\subsubsection{Limits on the CMF}
\label{sec:CMF}

So far we have only studied the detection probability of an individual cluster injected into the APOGEE data. For example, in the bottom panel of Fig.~\ref{fig:detection-proba}, we derived the probability of detecting a cluster as a function of its cluster mass, which we will denote as $P_{\rm detect} (M_{\rm cluster})$. In this section, we will propagate this individual cluster statistic to constrain the CMF. We derive the total number of clusters of different masses from $M_{\rm total}$ and the parameters of the CMF. Using this information, we can then evaluate the probability of all these predicted clusters being consistent with the APOGEE observation, which will then place a limit on the CMF. We assume a power-law CMF with a low-end cutoff of $30 \, {\rm M}_\odot$ and then constrain the power-law slope, $\alpha$, and the high-end cutoff, $M_{\rm cutoff}$ from the comparison with APOGEE data.

Let's formulate this idea more rigorously. Given a CMF, we know that, on average, there are a total of $\overline{n} = M_{\rm total}/\overline{M}$ clusters spawned where $\overline{M}$ is the mean cluster mass from the CMF. By definition, the cluster masses of $\overline{n}$ clusters follow the CMF, which we will denote as $M_{\rm cluster,i=1}, M_{\rm cluster,i=2}, \ldots, M_{\rm cluster,i=\overline{n}}$. The probability $\mathcal{L}({\rm CMF})$ that all these clusters are consistent with the data is
\begin{equation}
\label{eq:CMF-limit}
\mathcal{L}\big({\rm CMF}(\alpha, M_{\rm cutoff})\big) = \prod_{i=1}^{\overline{n}} \big(1-P_{\rm detect} (M_{\rm cluster,i})\big),
\end{equation}

\noindent
i.e., none of these clusters exceeds the observation baseline. In practice, to save computational time and to ensure a well-converged solution, we evaluate the $\mathcal{L}\big({\rm CMF})$ analytically.

The left panel of Fig.~\ref{fig:CMF-limit} shows the resulting $\mathcal{L}\big({\rm CMF})$. The figure demonstrates that if the CMF slope is shallower than $\alpha = -2$, the APOGEE sample is mostly consistent with a high-end cutoff $\gtrsim 3 \times 10^7 \, {\rm M}_\odot$ ($\log M_{\rm cutoff} = 7.5$). Qualitatively, this result should be expected. As shown in the bottom panel of Fig.~\ref{fig:detection-proba}, there is a $\sim 10 \%$ chance that a cluster with $\sim 10^7 \, {\rm M}_\odot$ will exceed the observation baseline. Recall that if the CMF slope $\alpha = -2$, we have equal contributions from all logarithmic mass bins. This implies that the number of clusters with mass $\sim 10^7 \, {\rm M}_\odot$ is of the order $\sim M_{\rm total}/10^7 \, {\rm M}_\odot \sim 100$. Let say there are $50$ such clusters, and each cluster only exceeds the baseline $\sim 10 \%$ of the time. The probability that all of them would be consistent with the APOGEE observation is still extremely unlikely because $(90\%)^{50} < 1\%$.  

This simple illustration also demonstrates two important features. First, the detection probability is very low for individual clusters with masses $<10^7 \, {\rm M}_\odot$. If the CMF slope is steeper than $-2$, most clusters are not massive. In this case, the APOGEE observation provides a very weak constraint on the high mass cutoff. As shown in the left panel of Fig.~\ref{fig:CMF-limit}, if $\alpha \lesssim -2$, we cannot constrain the CMF cutoff. Even though a cluster with mass $\sim 10^8 \, {\rm M}_\odot$ would easily exceed the baseline, these clusters are extremely rare if the CMF slope is steeper than $-2$.

Secondly, since the CMF constraint is derived from the product of each detection probability, it is sensitive to $P_{\rm detect}$. As shown in the bottom panel of Fig.~\ref{fig:detection-proba}, if our estimate of $\sigma_{\bf X}$ is off by $20\%$, it will affect $P_{\rm detect}$, which in turn could dramatically modify our CMF constraint. If clusters are more concentrated in abundance space than we have assumed here, then that will provide a stronger constraint on each detection (dashed line in Fig.~\ref{fig:detection-proba}). Therefore, our constraint on the CMF would be conservative. On the other hand, if we have underestimated $\sigma_{\bf X}$ by $20\%$, then the clusters would be more widely spread out in abundance space than we have assumed. In this case, most clusters would be harder to detect (dotted line in Fig.~\ref{fig:detection-proba}). Although not shown, we have checked that, in this case, we can only rule out CMF with $M_{\rm cutoff} \gtrsim 10^8 \, {\rm M}_\odot$ and $\alpha \gtrsim -1.9$.

Nonetheless, independent evidence seems to support our $\sigma_{\bf X}$ estimate. Mathematically, the rotation that we performed in \S\ref{sec:deconvolve} is exactly the same as the principal components analysis \citep[see appendix in][]{tin12a}. After the rotation, each coordinate becomes a principal component of the APOGEE abundance space. The APOGEE abundance space has fewer independent dimensions than the observed dimensions \citep{and12,tin12a}. Some of these 10 principal components should have very little intrinsic scatter. Therefore, some minor axes of the 10-dimensional ellipsoid are only due to the measurement scatter. Thus, their spreads should be a robust estimate of the measurement uncertainty $\sigma_{\bf X}$. We find that the widths of these minor axes are consistent with our $\sigma_{\bf X}$ estimate, showing our estimation of $\sigma_{\bf X}$ is robust. Therefore, our conservative CMF constraint is likely to hold.

Although not all $M_{\rm cutoff} \gtrsim 3 \times 10^7 \, {\rm M}_\odot$ is ruled out, Fig.~\ref{fig:CMF-limit} shows that a very high $M_{\rm cutoff}$ is only consistent with the data when the CMF slope is steeper than $\alpha = -2$. When the CMF is steep, the number of massive clusters also decreases precipitously. Therefore, the cutoff could be very massive, yet on average there might be less than one such massive cluster in the Milky Way. A high cutoff does not naturally imply the existence of these clusters. Instead of $M_{\rm cutoff}$, perhaps a more useful constraint is the maximum cluster mass such that we expect to have at least one cluster larger than this mass. We denote this maximum mass to be $M_{\rm max}$. Assuming $M_{\rm total} = 6 \times 10^9 \, {\rm M}_\odot$, we show $M_{\rm max}$ as a function of the CMF parameters in the right panel of Fig.~\ref{fig:CMF-limit}. As expected, this panels shows that when the slope is steep, we have $M_{\rm max} \ll M_{\rm cutoff}$, i.e., the cluster mass cutoff is never achieved. When overplotted with the constraints obtained in the left panel, the right panel shows that in most cases, only clusters with masses $\lesssim 3 \times 10^7$ could have formed. As the $\alpha$-enhanced disk is believed to form in the first 5 billion years \citep[e.g.,][]{hayw13}, our constraints refer to the portion of the disk that formed at $z > 1$.

We have made numerous assumptions in this study, but we emphasize that we have always made the conservative choices. Therefore, our CMF limit should be robust as long as we did not underestimate $\sigma_{\bf X}$ by $20\%$ and the ellipsoidal intrinsic distribution is a fair representation of the deconvolved distribution. A question remains to be answered: could we obtain a significantly stronger constraint on the CMF using the current APOGEE data? We would argue that the answer is likely no. The bottleneck is intrinsically due to the relatively small number of volume elements in abundance space and the low sampling rate.  The former is set by the precision of the abundance measurements and the number of independent dimensions in abundance space sampled by the APOGEE spectra. The APOGEE sampling rate is of the order $N_{\rm APOGEE}/M_{\rm total} = \mathcal{O}(10^{-5})$, which implies that we would only sample one star from a $\sim 10^5 \, {\rm M}_\odot$ cluster. The threshold $n_s > 10$ implies that the minimum cluster mass needed is $\sim 10^6 \, {\rm M}_\odot$. Therefore, in the most optimistic case, we might be able to put a stronger limit by at most an order of magnitude.

How does our stellar CMF limit compare to high redshift observations of star forming galaxies? Recent observations have reported the existence of giant star forming clumps within the disks of star forming galaxies at $z \sim 2$ \citep{gen06,for09,jon10,liv12,gen13}. Gas clumps as high as $\sim 10^9 \, {\rm M}_\odot$ have been observed.  There has been much speculation regarding the properties and fate of these giant clumps.  Some have argued that they are progenitors of globular clusters \citep{sha10}.  They may migrate by dynamical friction to the centers of galaxies \citep{wuy12}.  It is also unclear if the stars forming within these giant clumps contain stars that are coeval or share a common metallicity.  What is clear from the results presented in this work is that these giant star forming clumps cannot both be mono-abundance and remain in the $\alpha$-enhanced disk, at least in the portion of the Milky Way observed by APOGEE (i.e., with a Galactocentric radius of $3\,{\rm kpc}-15\,{\rm kpc}$). Even assuming a total star formation efficiency of $1\%$ \citep[simulations and observations usually show higher values, e.g.][]{ken98,elm02,kru05,eva09,krui12}, these gas clumps would have formed clusters that are at least $10^7 \, {\rm M}_\odot$ and would have stood out in the chemical tagging experiment presented here if they are chemically homogeneous and that they remain in the Milky Way disk, which APOGEE is probing.

%
%
%
%
%
%

\subsection{Comparison with previous studies}
\label{sec:claims}

The first chemical tagging experiment on dispersed disk stars was performed by \citet{mit14}, and the tagged groups were subsequently studied in \citet{qui15}. They studied 714 stars in the solar neighborhood from \citet{ben14}. Our results agree with their assessments that the identified groups in these studies are probably not co-natal stars. Each group is unlikely coming from a single disrupted cluster, even though the clump members might be coeval stars as they share similar abundances. In these studies, the sample includes both $\alpha$-enhanced stars and low-$\alpha$ stars, but the sample size is about ten times smaller than the APOGEE $\alpha$-enhanced sample. The sampling rate in \citet{ben14} is therefore much smaller than the APOGEE $\alpha$-enhanced sample. Recall that the sampling rate in this study is $\sim \mathcal{O}(10^{-5})$, and thus, we deduce that the sampling rate in these early studies is $\ll \mathcal{O} (10^{-5})/10 = \mathcal{O} (10^{-6})$. If groups detected in these studies were to come from individually disrupted clusters, the parent cluster would have a mass $\gg 10^6 {\rm M}_{\odot}$, consistent with the estimates in \citet{qui15}.

Simulations from \citet{tin15} also disfavor a co-natal interpretation of the groups identified in these earlier studies. \citet{tin15} found that even if such large clusters exist, a detected clump in abundance space will still have a sizable background component. More importantly, in the case with a dominant background, the applicability of previous clump search techniques that separate the abundance space into a few distinct regions, such as the one proposed in \citet{mit14}, or other tree-based methods \citep[e.g.][]{mac15} is questionable. For those techniques to perform well, the background in abundance space has to be negligible or first be subtracted. As we have explored in this study, the background estimation can be challenging given its complex topology in high-dimensional space and the fact that the signal is usually overwhelmed by the background contaminants.

%
%
%
%
%
%

\section{Summary and conclusion}
\label{sec:conclusion}

In this study we have exploited the superb APOGEE DR12 data, with typical uncertainties of $\sigma_{[X/{\rm Fe}]}$ and $\sigma_{\rm [Fe/H]} \simeq 0.05$ for 10 elements measured for $>10^4$ stars, to put in practice a first large-scale chemical tagging analysis of the $\alpha$-enhanced disk. Because the number of stars per 10-dimensional abundance volume is lower in the $\alpha$-enhanced disk, we focused on that portion of abundance space.

This analysis required the development of a new, simple algorithm for identifying clumps in abundance space, and it brought some of the ``real life'' difficulties of chemical tagging to the fore. Nonetheless, we succeeded in providing the first abundance-based constraints on the masses and mass functions of chemically homogeneous star clusters in the old Galactic disk.

The methodological steps and results can be summarized as follows:

\begin{itemize}

\item We determined and applied a coordinate transformation that makes the cluster kernel in abundance space spherical (in 10-dimensions) and have unit variance in each dimension. This kernel enables fast error deconvolutions in this transformed abundance space. 

\item We generated a model for the intrinsic abundance distribution of the $\alpha$-enhanced disk, presuming it to be a highly anisotropic and co-variant ellipsoidal distribution in the above 10-dimensional transformed abundance space. After rotating this coordinate system to eliminate the co-variances in this distribution, we modeled each dimension independently as the sum of two Gaussians. Fitting this to the APOGEE data provides a first estimate of the shape and volume of the error-deconvolved abundance space of the $\alpha$-enhanced Galactic disk. 

\item We found that despite the unprecedented quality of the APOGEE data, the volume occupied by the stars of the $\alpha$-enhanced Galactic disk is only $\sim 500$ times the volume of the cluster kernel. Even with abundance uncertainties of $\sigma_{[X/{\rm Fe}]}$ and $\sigma_{\rm [Fe/H]} \simeq 0.05 \, {\rm dex}$, the cluster kernel spans $> 30\%$ of the abundance width in each element abundance dimension. In addition, many of the 10 element abundances measured by APOGEE and used herein are highly co-variant.

\item We developed an algorithm to detect groups of chemically homogeneous stars, geared toward the background-dominated regime. We found that searching for chemically homogeneous clumps is challenging
with high backgrounds. The chemical tagging signals will most likely come from the peripheral regions in abundance space where the background density is relatively low.

\item Using APOGEE data as a detection baseline, we were able to constrain the CMF in the Galactic $\alpha$-enhanced disk. We show that this population is unlikely to have formed clusters more massive than $3 \times 10^7 \, {\rm M}_\odot$ at any point in its history.
\end{itemize}

Although the current constraints presented in this work are limited to very large cluster masses, the results in this paper vividly demonstrate the potential of chemical tagging in understanding the Milky Way properties in the past. With more data currently being collected by on-going surveys, we should be able to provide much stronger constraints on the CMF in the near future.

%
%
%
%
%
%

\acknowledgments

We thank the referees for the careful and insightful comments. This research was inspired at the KITP Galatic Archaeology and Precision Stellar Astrophysics workshop in year 2015 and was supported in part by the National Science Foundation under Grant No. NSF PHY11-25915. YST is grateful to the Max-Planck-Institut f\"{u}r Astronomie and the DFG through the SFB 881 (A3) for their hospitality and financial support during the period in which part of this research was performed. HWR's research contribution is supported by the European Research Council under the European Union's Seventh Framework Programme (FP 7) ERC Grant Agreement n. [321035].

%
%
%
%
%
%

\end{CJK*}

\bibliography{biblio.bib}

\begin{thebibliography}{}
\expandafter\ifx\csname natexlab\endcsname\relax\def\natexlab#1{#1}\fi

\bibitem[{{Adamo} {et~al.}(2015){Adamo}, {Kruijssen}, {Bastian}, {Silva-Villa},
  \& {Ryon}}]{ada15}
{Adamo}, A., {Kruijssen}, J.~M.~D., {Bastian}, N., {Silva-Villa}, E., \&
  {Ryon}, J. 2015, \mnras, 452, 246

\bibitem[{{Andrews} {et~al.}(2012){Andrews}, {Weinberg}, {Johnson}, {Bensby},
  \& {Feltzing}}]{and12}
{Andrews}, B.~H., {Weinberg}, D.~H., {Johnson}, J.~A., {Bensby}, T., \&
  {Feltzing}, S. 2012, AcA, 62, 269

\bibitem[{{Barklem} {et~al.}(2005){Barklem}, {Christlieb}, {Beers}, {Hill},
  {Bessell}, {Holmberg}, {Marsteller}, {Rossi}, {Zickgraf}, \&
  {Reimers}}]{bar05}
{Barklem}, P.~S., {Christlieb}, N., {Beers}, T.~C., {et~al.} 2005, \aap, 439,
  129

\bibitem[{{Bensby} {et~al.}(2014){Bensby}, {Feltzing}, \& {Oey}}]{ben14}
{Bensby}, T., {Feltzing}, S., \& {Oey}, M.~S. 2014, \aap, 562, A71

\bibitem[{{Bica} {et~al.}(2003){Bica}, {Dutra}, {Soares}, \& {Barbuy}}]{bic03}
{Bica}, E., {Dutra}, C.~M., {Soares}, J., \& {Barbuy}, B. 2003, \aap, 404, 223

\bibitem[{{Bird} {et~al.}(2013){Bird}, {Kazantzidis}, {Weinberg}, {Guedes},
  {Callegari}, {Mayer}, \& {Madau}}]{bir13}
{Bird}, J.~C., {Kazantzidis}, S., {Weinberg}, D.~H., {et~al.} 2013, \apj, 773,
  43

\bibitem[{{Blanco-Cuaresma} {et~al.}(2015){Blanco-Cuaresma}, {Soubiran},
  {Heiter}, {Asplund}, {Carraro}, {Costado}, {Feltzing},
  {Gonz{\'a}lez-Hern{\'a}ndez}, {Jim{\'e}nez-Esteban}, {Korn}, {Marino},
  {Montes}, {San Roman}, {Tabernero}, \& {Tautvai{\v s}ien{\.e}}}]{blan15}
{Blanco-Cuaresma}, S., {Soubiran}, C., {Heiter}, U., {et~al.} 2015, \aap, 577,
  A47

\bibitem[{{Bland-Hawthorn} {et~al.}(2010a){Bland-Hawthorn}, {Karlsson},
  {Sharma}, {Krumholz}, \& {Silk}}]{bla10a}
{Bland-Hawthorn}, J., {Karlsson}, T., {Sharma}, S., {Krumholz}, M., \& {Silk},
  J. 2010a, \apj, 721, 582

\bibitem[{{Bland-Hawthorn} {et~al.}(2010b){Bland-Hawthorn}, {Krumholz}, \&
  {Freeman}}]{bla10b}
{Bland-Hawthorn}, J., {Krumholz}, M.~R., \& {Freeman}, K. 2010b, \apj, 713, 166

\bibitem[{{Bland-Hawthorn} {et~al.}(2014){Bland-Hawthorn}, {Sharma}, \&
  {Freeman}}]{bla15}
{Bland-Hawthorn}, J., {Sharma}, S., \& {Freeman}, K. 2014, in EAS Publications
  Series, Vol.~67, EAS Publications Series, 219--226

\bibitem[{{Borissova} {et~al.}(2011){Borissova}, {Bonatto}, {Kurtev}, {Clarke},
  {Pe{\~n}aloza}, {Sale}, {Minniti}, {Alonso-Garc{\'{\i}}a}, {Artigau},
  {Barb{\'a}}, {Bica}, {Baume}, {Catelan}, {Chen{\`e}}, {Dias}, {Folkes},
  {Froebrich}, {Geisler}, {de Grijs}, {Hanson}, {Hempel}, {Ivanov}, {Kumar},
  {Lucas}, {Mauro}, {Moni Bidin}, {Rejkuba}, {Saito}, {Tamura}, \&
  {Toledo}}]{bor11}
{Borissova}, J., {Bonatto}, C., {Kurtev}, R., {et~al.} 2011, \aap, 532, A131

\bibitem[{{Bournaud} {et~al.}(2009){Bournaud}, {Elmegreen}, \&
  {Martig}}]{bou09}
{Bournaud}, F., {Elmegreen}, B.~G., \& {Martig}, M. 2009, \apjl, 707, L1

\bibitem[{{Bovy} \& {Rix}(2013)}]{bov13}
{Bovy}, J., \& {Rix}, H.-W. 2013, \apj, 779, 115

\bibitem[{{Bovy} {et~al.}(2012){Bovy}, {Rix}, {Liu}, {Hogg}, {Beers}, \&
  {Lee}}]{bov12}
{Bovy}, J., {Rix}, H.-W., {Liu}, C., {et~al.} 2012, \apj, 753, 148

\bibitem[{{Bovy} {et~al.}(2014){Bovy}, {Nidever}, {Rix}, {Girardi}, {Zasowski},
  {Chojnowski}, {Holtzman}, {Epstein}, {Frinchaboy}, {Hayden}, {Rodrigues},
  {Majewski}, {Johnson}, {Pinsonneault}, {Stello}, {Allende Prieto}, {Andrews},
  {Basu}, {Beers}, {Bizyaev}, {Burton}, {Chaplin}, {Cunha}, {Elsworth},
  {Garc{\'{\i}}a}, {Garc{\'{\i}}a-Her{\'n}andez}, {Garc{\'{\i}}a P{\'e}rez},
  {Hearty}, {Hekker}, {Kallinger}, {Kinemuchi}, {Koesterke},
  {M{\'e}sz{\'a}ros}, {Mosser}, {O'Connell}, {Oravetz}, {Pan}, {Robin},
  {Schiavon}, {Schneider}, {Schultheis}, {Serenelli}, {Shetrone}, {Silva
  Aguirre}, {Simmons}, {Skrutskie}, {Smith}, {Stassun}, {Weinberg}, {Wilson},
  \& {Zamora}}]{bov14}
{Bovy}, J., {Nidever}, D.~L., {Rix}, H.-W., {et~al.} 2014, \apj, 790, 127

\bibitem[{{Bragaglia} {et~al.}(2012){Bragaglia}, {Gratton}, {Carretta},
  {D'Orazi}, {Sneden}, \& {Lucatello}}]{bra12}
{Bragaglia}, A., {Gratton}, R.~G., {Carretta}, E., {et~al.} 2012, \aap, 548,
  A122

\bibitem[{{Brandner} {et~al.}(2008){Brandner}, {Clark}, {Stolte}, {Waters},
  {Negueruela}, \& {Goodwin}}]{bra08}
{Brandner}, W., {Clark}, J.~S., {Stolte}, A., {et~al.} 2008, \aap, 478, 137

\bibitem[{{Bubar} \& {King}(2010)}]{bub10}
{Bubar}, E.~J., \& {King}, J.~R. 2010, \aj, 140, 293

\bibitem[{{Cheng} {et~al.}(2012){Cheng}, {Rockosi}, {Morrison}, {Lee}, {Beers},
  {Bizyaev}, {Harding}, {Malanushenko}, {Malanushenko}, {Oravetz}, {Pan},
  {Schlesinger}, {Schneider}, {Simmons}, \& {Weaver}}]{che12}
{Cheng}, J.~Y., {Rockosi}, C.~M., {Morrison}, H.~L., {et~al.} 2012, \apj, 752,
  51

\bibitem[{{Conroy} {et~al.}(2009){Conroy}, {Gunn}, \& {White}}]{con09}
{Conroy}, C., {Gunn}, J.~E., \& {White}, M. 2009, \apj, 699, 486

\bibitem[{{Dalessandro} {et~al.}(2015){Dalessandro}, {Miocchi}, {Carraro},
  {J{\'{\i}}lkov{\'a}}, \& {Moitinho}}]{dal15}
{Dalessandro}, E., {Miocchi}, P., {Carraro}, G., {J{\'{\i}}lkov{\'a}}, L., \&
  {Moitinho}, A. 2015, \mnras, 449, 1811

\bibitem[{{De Silva} {et~al.}(2013){De Silva}, {D'Orazi}, {Melo}, {Torres},
  {Gieles}, {Quast}, \& {Sterzik}}]{des13}
{De Silva}, G.~M., {D'Orazi}, V., {Melo}, C., {et~al.} 2013, \mnras, 431, 1005

\bibitem[{{De Silva} {et~al.}(2007b){De Silva}, {Freeman}, {Asplund},
  {Bland-Hawthorn}, {Bessell}, \& {Collet}}]{des07b}
{De Silva}, G.~M., {Freeman}, K.~C., {Asplund}, M., {et~al.} 2007b, \aj, 133,
  1161

\bibitem[{{De Silva} {et~al.}(2009){De Silva}, {Freeman}, \&
  {Bland-Hawthorn}}]{des09}
{De Silva}, G.~M., {Freeman}, K.~C., \& {Bland-Hawthorn}, J. 2009, \pasa, 26,
  11

\bibitem[{{De Silva} {et~al.}(2007a){De Silva}, {Freeman}, {Bland-Hawthorn},
  {Asplund}, \& {Bessell}}]{des07a}
{De Silva}, G.~M., {Freeman}, K.~C., {Bland-Hawthorn}, J., {Asplund}, M., \&
  {Bessell}, M.~S. 2007a, \aj, 133, 694

\bibitem[{{De Silva} {et~al.}(2015){De Silva}, {Freeman}, {Bland-Hawthorn},
  {Martell}, {de Boer}, {Asplund}, {Keller}, {Sharma}, {Zucker}, {Zwitter},
  {Anguiano}, {Bacigalupo}, {Bayliss}, {Beavis}, {Bergemann}, {Campbell},
  {Cannon}, {Carollo}, {Casagrande}, {Casey}, {Da Costa}, {D'Orazi}, {Dotter},
  {Duong}, {Heger}, {Ireland}, {Kafle}, {Kos}, {Lattanzio}, {Lewis}, {Lin},
  {Lind}, {Munari}, {Nataf}, {O'Toole}, {Parker}, {Reid}, {Schlesinger},
  {Sheinis}, {Simpson}, {Stello}, {Ting}, {Traven}, {Watson}, {Wittenmyer},
  {Yong}, \& {{\v Z}erjal}}]{des15}
{De Silva}, G.~M., {Freeman}, K.~C., {Bland-Hawthorn}, J., {et~al.} 2015,
  \mnras, 449, 2604

\bibitem[{{Di Matteo} {et~al.}(2013){Di Matteo}, {Haywood}, {Combes},
  {Semelin}, \& {Snaith}}]{dim13}
{Di Matteo}, P., {Haywood}, M., {Combes}, F., {Semelin}, B., \& {Snaith}, O.~N.
  2013, \aap, 553, A102

\bibitem[{{Edvardsson} {et~al.}(1993){Edvardsson}, {Andersen}, {Gustafsson},
  {Lambert}, {Nissen}, \& {Tomkin}}]{edv93}
{Edvardsson}, B., {Andersen}, J., {Gustafsson}, B., {et~al.} 1993, \aap, 275,
  101

\bibitem[{{Elmegreen}(2002)}]{elm02}
{Elmegreen}, B.~G. 2002, \apj, 577, 206

\bibitem[{{Escala} \& {Larson}(2008)}]{esc08}
{Escala}, A., \& {Larson}, R.~B. 2008, \apjl, 685, L31

\bibitem[{{Evans} {et~al.}(2009){Evans}, {Dunham}, {J{\o}rgensen}, {Enoch},
  {Mer{\'{\i}}n}, {van Dishoeck}, {Alcal{\'a}}, {Myers}, {Stapelfeldt},
  {Huard}, {Allen}, {Harvey}, {van Kempen}, {Blake}, {Koerner}, {Mundy},
  {Padgett}, \& {Sargent}}]{eva09}
{Evans}, II, N.~J., {Dunham}, M.~M., {J{\o}rgensen}, J.~K., {et~al.} 2009,
  \apjs, 181, 321

\bibitem[{{Everitt} {et~al.}(2010){Everitt}, {Landau}, {Leese}, \&
  {Stahl}}]{eve10}
{Everitt}, B.~S., {Landau}, S., {Leese}, M., \& {Stahl}, D. 2010, {Cluster
  Analysis: 5th Edition} (John Wiley \& Sons)

\bibitem[{{Feng} \& {Krumholz}(2014)}]{fen14}
{Feng}, Y., \& {Krumholz}, M.~R. 2014, \nat, 513, 523

\bibitem[{{Flynn} {et~al.}(2006){Flynn}, {Holmberg}, {Portinari}, {Fuchs}, \&
  {Jahrei{\ss}}}]{fly06}
{Flynn}, C., {Holmberg}, J., {Portinari}, L., {Fuchs}, B., \& {Jahrei{\ss}}, H.
  2006, \mnras, 372, 1149

\bibitem[{{F{\"o}rster Schreiber} {et~al.}(2009){F{\"o}rster Schreiber},
  {Genzel}, {Bouch{\'e}}, {Cresci}, {Davies}, {Buschkamp}, {Shapiro},
  {Tacconi}, {Hicks}, {Genel}, {Shapley}, {Erb}, {Steidel}, {Lutz},
  {Eisenhauer}, {Gillessen}, {Sternberg}, {Renzini}, {Cimatti}, {Daddi},
  {Kurk}, {Lilly}, {Kong}, {Lehnert}, {Nesvadba}, {Verma}, {McCracken},
  {Arimoto}, {Mignoli}, \& {Onodera}}]{for09}
{F{\"o}rster Schreiber}, N.~M., {Genzel}, R., {Bouch{\'e}}, N., {et~al.} 2009,
  \apj, 706, 1364

\bibitem[{{Freeman} \& {Bland-Hawthorn}(2002)}]{fre02}
{Freeman}, K., \& {Bland-Hawthorn}, J. 2002, \araa, 40, 487

\bibitem[{{Friel} {et~al.}(2014){Friel}, {Donati}, {Bragaglia}, {Jacobson},
  {Magrini}, {Prisinzano}, {Randich}, {Tosi}, {Cantat-Gaudin}, {Vallenari},
  {Smiljanic}, {Carraro}, {Sordo}, {Maiorca}, {Tautvai{\v s}ien{\.e}},
  {Sestito}, {Zaggia}, {Jim{\'e}nez-Esteban}, {Gilmore}, {Jeffries}, {Alfaro},
  {Bensby}, {Koposov}, {Korn}, {Pancino}, {Recio-Blanco}, {Franciosini},
  {Hill}, {Jackson}, {de Laverny}, {Morbidelli}, {Sacco}, {Worley},
  {Hourihane}, {Costado}, {Jofr{\'e}}, \& {Lind}}]{fri14}
{Friel}, E.~D., {Donati}, P., {Bragaglia}, A., {et~al.} 2014, \aap, 563, A117

\bibitem[{{Genzel} {et~al.}(2006){Genzel}, {Tacconi}, {Eisenhauer},
  {F{\"o}rster Schreiber}, {Cimatti}, {Daddi}, {Bouch{\'e}}, {Davies},
  {Lehnert}, {Lutz}, {Nesvadba}, {Verma}, {Abuter}, {Shapiro}, {Sternberg},
  {Renzini}, {Kong}, {Arimoto}, \& {Mignoli}}]{gen06}
{Genzel}, R., {Tacconi}, L.~J., {Eisenhauer}, F., {et~al.} 2006, \nat, 442, 786

\bibitem[{{Genzel} {et~al.}(2013){Genzel}, {Tacconi}, {Kurk}, {Wuyts},
  {Combes}, {Freundlich}, {Bolatto}, {Cooper}, {Neri}, {Nordon}, {Bournaud},
  {Burkert}, {Comerford}, {Cox}, {Davis}, {F{\"o}rster Schreiber},
  {Garc{\'{\i}}a-Burillo}, {Gracia-Carpio}, {Lutz}, {Naab}, {Newman},
  {Saintonge}, {Shapiro Griffin}, {Shapley}, {Sternberg}, \& {Weiner}}]{gen13}
{Genzel}, R., {Tacconi}, L.~J., {Kurk}, J., {et~al.} 2013, \apj, 773, 68

\bibitem[{{Ghez} {et~al.}(2008){Ghez}, {Salim}, {Weinberg}, {Lu}, {Do}, {Dunn},
  {Matthews}, {Morris}, {Yelda}, {Becklin}, {Kremenek}, {Milosavljevic}, \&
  {Naiman}}]{ghe08}
{Ghez}, A.~M., {Salim}, S., {Weinberg}, N.~N., {et~al.} 2008, \apj, 689, 1044

\bibitem[{{Gillessen} {et~al.}(2009){Gillessen}, {Eisenhauer}, {Trippe},
  {Alexander}, {Genzel}, {Martins}, \& {Ott}}]{gil09}
{Gillessen}, S., {Eisenhauer}, F., {Trippe}, S., {et~al.} 2009, \apj, 692, 1075

\bibitem[{{Gould} \& {Rix}(2015)}]{gou15}
{Gould}, A., \& {Rix}, H.-W. 2015, ArXiv e-prints, submitted to JKAS,
  arXiv:1502.05709

\bibitem[{{Halle} {et~al.}(2015){Halle}, {Di Matteo}, {Haywood}, \&
  {Combes}}]{hal15}
{Halle}, A., {Di Matteo}, P., {Haywood}, M., \& {Combes}, F. 2015, \aap, 578,
  A58

\bibitem[{{Hayden} {et~al.}(2015){Hayden}, {Bovy}, {Holtzman}, {Nidever},
  {Bird}, {Weinberg}, {Andrews}, {Majewski}, {Allende Prieto}, {Anders},
  {Beers}, {Bizyaev}, {Chiappini}, {Cunha}, {Frinchaboy},
  {Garc{\'{\i}}a-Her{\'n}andez}, {Garc{\'{\i}}a P{\'e}rez}, {Girardi},
  {Harding}, {Hearty}, {Johnson}, {M{\'e}sz{\'a}ros}, {Minchev}, {O'Connell},
  {Pan}, {Robin}, {Schiavon}, {Schneider}, {Schultheis}, {Shetrone},
  {Skrutskie}, {Steinmetz}, {Smith}, {Wilson}, {Zamora}, \& {Zasowski}}]{hay15}
{Hayden}, M.~R., {Bovy}, J., {Holtzman}, J.~A., {et~al.} 2015, \apj, 808, 132

\bibitem[{{Haywood}(2008)}]{hayw08}
{Haywood}, M. 2008, \mnras, 388, 1175

\bibitem[{{Haywood} {et~al.}(2013){Haywood}, {Di Matteo}, {Lehnert}, {Katz}, \&
  {G{\'o}mez}}]{hayw13}
{Haywood}, M., {Di Matteo}, P., {Lehnert}, M.~D., {Katz}, D., \& {G{\'o}mez},
  A. 2013, \aap, 560, A109

\bibitem[{{Holtzman} {et~al.}(2015){Holtzman}, {Shetrone}, {Johnson}, {Allende
  Prieto}, {Anders}, {Andrews}, {Beers}, {Bizyaev}, {Blanton}, {Bovy},
  {Carrera}, {Chojnowski}, {Cunha}, {Eisenstein}, {Feuillet}, {Frinchaboy},
  {Galbraith-Frew}, {Garc{\'{\i}}a P{\'e}rez}, {Garc{\'{\i}}a-Hern{\'a}ndez},
  {Hasselquist}, {Hayden}, {Hearty}, {Ivans}, {Majewski}, {Martell},
  {Meszaros}, {Muna}, {Nidever}, {Nguyen}, {O'Connell}, {Pan}, {Pinsonneault},
  {Robin}, {Schiavon}, {Shane}, {Sobeck}, {Smith}, {Troup}, {Weinberg},
  {Wilson}, {Wood-Vasey}, {Zamora}, \& {Zasowski}}]{hol15}
{Holtzman}, J.~A., {Shetrone}, M., {Johnson}, J.~A., {et~al.} 2015, \aj, 150,
  148

\bibitem[{{Jones} {et~al.}(2010){Jones}, {Swinbank}, {Ellis}, {Richard}, \&
  {Stark}}]{jon10}
{Jones}, T.~A., {Swinbank}, A.~M., {Ellis}, R.~S., {Richard}, J., \& {Stark},
  D.~P. 2010, \mnras, 404, 1247

\bibitem[{{Karakas} \& {Lattanzio}(2014)}]{kar14}
{Karakas}, A.~I., \& {Lattanzio}, J.~C. 2014, \pasa, 31, 30

\bibitem[{{Karlsson} {et~al.}(2012){Karlsson}, {Bland-Hawthorn}, {Freeman}, \&
  {Silk}}]{kar12}
{Karlsson}, T., {Bland-Hawthorn}, J., {Freeman}, K.~C., \& {Silk}, J. 2012,
  \apj, 759, 111

\bibitem[{{Kennicutt}(1998)}]{ken98}
{Kennicutt}, Jr., R.~C. 1998, \apj, 498, 541

\bibitem[{{Kobayashi} {et~al.}(2011){Kobayashi}, {Karakas}, \& {Umeda}}]{kob11}
{Kobayashi}, C., {Karakas}, A.~I., \& {Umeda}, H. 2011, \mnras, 414, 3231

\bibitem[{{Kobayashi} {et~al.}(2006){Kobayashi}, {Umeda}, {Nomoto}, {Tominaga},
  \& {Ohkubo}}]{kob06}
{Kobayashi}, C., {Umeda}, H., {Nomoto}, K., {Tominaga}, N., \& {Ohkubo}, T.
  2006, \apj, 653, 1145

\bibitem[{{Koposov} {et~al.}(2008){Koposov}, {Glushkova}, \&
  {Zolotukhin}}]{kop08}
{Koposov}, S.~E., {Glushkova}, E.~V., \& {Zolotukhin}, I.~Y. 2008, \aap, 486,
  771

\bibitem[{{Koposov} {et~al.}(2010){Koposov}, {Rix}, \& {Hogg}}]{kop10}
{Koposov}, S.~E., {Rix}, H.-W., \& {Hogg}, D.~W. 2010, \apj, 712, 260

\bibitem[{{Kordopatis} {et~al.}(2015){Kordopatis}, {Binney}, {Gilmore}, {Wyse},
  {Belokurov}, {McMillan}, {Hatfield}, {Grebel}, {Steinmetz}, {Navarro},
  {Seabroke}, {Minchev}, {Chiappini}, {Bienaym{\'e}}, {Bland-Hawthorn},
  {Freeman}, {Gibson}, {Helmi}, {Munari}, {Parker}, {Reid}, {Siebert},
  {Siviero}, \& {Zwitter}}]{kor15}
{Kordopatis}, G., {Binney}, J., {Gilmore}, G., {et~al.} 2015, \mnras, 447, 3526

\bibitem[{{Kruijssen}(2012)}]{krui12}
{Kruijssen}, J.~M.~D. 2012, \mnras, 426, 3008

\bibitem[{{Krumholz} {et~al.}(2012){Krumholz}, {Dekel}, \& {McKee}}]{kru12}
{Krumholz}, M.~R., {Dekel}, A., \& {McKee}, C.~F. 2012, \apj, 745, 69

\bibitem[{{Krumholz} \& {McKee}(2005)}]{kru05}
{Krumholz}, M.~R., \& {McKee}, C.~F. 2005, \apj, 630, 250

\bibitem[{{Lada} \& {Lada}(2003)}]{lad03}
{Lada}, C.~J., \& {Lada}, E.~A. 2003, \araa, 41, 57

\bibitem[{{Livermore} {et~al.}(2012){Livermore}, {Jones}, {Richard}, {Bower},
  {Ellis}, {Swinbank}, {Rigby}, {Smail}, {Arribas}, {Rodriguez Zaurin},
  {Colina}, {Ebeling}, \& {Crain}}]{liv12}
{Livermore}, R.~C., {Jones}, T., {Richard}, J., {et~al.} 2012, \mnras, 427, 688

\bibitem[{{Loebman} {et~al.}(2011){Loebman}, {Ro{\v s}kar}, {Debattista},
  {Ivezi{\'c}}, {Quinn}, \& {Wadsley}}]{loe11}
{Loebman}, S.~R., {Ro{\v s}kar}, R., {Debattista}, V.~P., {et~al.} 2011, \apj,
  737, 8

\bibitem[{{Macfarlane} {et~al.}(2015){Macfarlane}, {Gibson}, \&
  {Flynn}}]{mac15}
{Macfarlane}, B.~A., {Gibson}, B.~K., \& {Flynn}, C.~M.~L. 2015, ArXiv
  e-prints, arXiv:1505.02059

\bibitem[{{Martig} {et~al.}(2015){Martig}, {Rix}, {Aguirre}, {Hekker},
  {Mosser}, {Elsworth}, {Bovy}, {Stello}, {Anders}, {Garc{\'{\i}}a}, {Tayar},
  {Rodrigues}, {Basu}, {Carrera}, {Ceillier}, {Chaplin}, {Chiappini},
  {Frinchaboy}, {Garc{\'{\i}}a-Hern{\'a}ndez}, {Hearty}, {Holtzman}, {Johnson},
  {Majewski}, {Mathur}, {M{\'e}sz{\'a}ros}, {Miglio}, {Nidever}, {Pan},
  {Pinsonneault}, {Schiavon}, {Schneider}, {Serenelli}, {Shetrone}, \&
  {Zamora}}]{mar15}
{Martig}, M., {Rix}, H.-W., {Aguirre}, V.~S., {et~al.} 2015, \mnras, 451, 2230

\bibitem[{{Masseron} \& {Gilmore}(2015)}]{mas15}
{Masseron}, T., \& {Gilmore}, G. 2015, \mnras, 453, 1855

\bibitem[{{McLachlan} \& {Peel}(2000)}]{mcl00}
{McLachlan}, G., \& {Peel}, D. 2000, {Finite Mixture Models: 1st Edition} (John
  Wiley \& Sons)

\bibitem[{{M{\'e}sz{\'a}ros} {et~al.}(2013){M{\'e}sz{\'a}ros}, {Holtzman},
  {Garc{\'{\i}}a P{\'e}rez}, {Allende Prieto}, {Schiavon}, {Basu}, {Bizyaev},
  {Chaplin}, {Chojnowski}, {Cunha}, {Elsworth}, {Epstein}, {Frinchaboy},
  {Garc{\'{\i}}a}, {Hearty}, {Hekker}, {Johnson}, {Kallinger}, {Koesterke},
  {Majewski}, {Martell}, {Nidever}, {Pinsonneault}, {O'Connell}, {Shetrone},
  {Smith}, {Wilson}, \& {Zasowski}}]{mes13}
{M{\'e}sz{\'a}ros}, S., {Holtzman}, J., {Garc{\'{\i}}a P{\'e}rez}, A.~E.,
  {et~al.} 2013, \aj, 146, 133

\bibitem[{{Minchev} {et~al.}(2013){Minchev}, {Chiappini}, \& {Martig}}]{min13}
{Minchev}, I., {Chiappini}, C., \& {Martig}, M. 2013, \aap, 558, A9

\bibitem[{{Minchev} \& {Famaey}(2010)}]{min10}
{Minchev}, I., \& {Famaey}, B. 2010, \apj, 722, 112

\bibitem[{{Mitschang} {et~al.}(2013){Mitschang}, {De Silva}, {Sharma}, \&
  {Zucker}}]{mit13}
{Mitschang}, A.~W., {De Silva}, G., {Sharma}, S., \& {Zucker}, D.~B. 2013,
  \mnras, 428, 2321

\bibitem[{{Mitschang} {et~al.}(2014){Mitschang}, {De Silva}, {Zucker},
  {Anguiano}, {Bensby}, \& {Feltzing}}]{mit14}
{Mitschang}, A.~W., {De Silva}, G., {Zucker}, D.~B., {et~al.} 2014, \mnras,
  438, 2753

\bibitem[{{Nidever} {et~al.}(2014){Nidever}, {Bovy}, {Bird}, {Andrews},
  {Hayden}, {Holtzman}, {Majewski}, {Smith}, {Robin}, {Garc{\'{\i}}a
  P{\'e}rez}, {Cunha}, {Allende Prieto}, {Zasowski}, {Schiavon}, {Johnson},
  {Weinberg}, {Feuillet}, {Schneider}, {Shetrone}, {Sobeck},
  {Garc{\'{\i}}a-Hern{\'a}ndez}, {Zamora}, {Rix}, {Beers}, {Wilson},
  {O'Connell}, {Minchev}, {Chiappini}, {Anders}, {Bizyaev}, {Brewington},
  {Ebelke}, {Frinchaboy}, {Ge}, {Kinemuchi}, {Malanushenko}, {Malanushenko},
  {Marchante}, {M{\'e}sz{\'a}ros}, {Oravetz}, {Pan}, {Simmons}, \&
  {Skrutskie}}]{nid14}
{Nidever}, D.~L., {Bovy}, J., {Bird}, J.~C., {et~al.} 2014, \apj, 796, 38

\bibitem[{{Odenkirchen} {et~al.}(2003){Odenkirchen}, {Grebel}, {Dehnen}, {Rix},
  {Yanny}, {Newberg}, {Rockosi}, {Mart{\'{\i}}nez-Delgado}, {Brinkmann}, \&
  {Pier}}]{ode03}
{Odenkirchen}, M., {Grebel}, E.~K., {Dehnen}, W., {et~al.} 2003, \aj, 126, 2385

\bibitem[{{Porras} {et~al.}(2003){Porras}, {Christopher}, {Allen}, {Di
  Francesco}, {Megeath}, \& {Myers}}]{por03}
{Porras}, A., {Christopher}, M., {Allen}, L., {et~al.} 2003, \aj, 126, 1916

\bibitem[{{Quillen} {et~al.}(2015){Quillen}, {Anguiano}, {De Silva}, {Freeman},
  {Zucker}, {Minchev}, \& {Bland-Hawthorn}}]{qui15}
{Quillen}, A.~C., {Anguiano}, B., {De Silva}, G., {et~al.} 2015, \mnras, 450,
  2354

\bibitem[{{Randich} {et~al.}(2013){Randich}, {Gilmore}, \& {Gaia-ESO
  Consortium}}]{ran13}
{Randich}, S., {Gilmore}, G., \& {Gaia-ESO Consortium}. 2013, The Messenger,
  154, 47

\bibitem[{{Reddy} {et~al.}(2006){Reddy}, {Lambert}, \& {Allende
  Prieto}}]{red06}
{Reddy}, B.~E., {Lambert}, D.~L., \& {Allende Prieto}, C. 2006, \mnras, 367,
  1329

\bibitem[{{Reid} {et~al.}(2014){Reid}, {Menten}, {Brunthaler}, {Zheng}, {Dame},
  {Xu}, {Wu}, {Zhang}, {Sanna}, {Sato}, {Hachisuka}, {Choi}, {Immer},
  {Moscadelli}, {Rygl}, \& {Bartkiewicz}}]{rei14}
{Reid}, M.~J., {Menten}, K.~M., {Brunthaler}, A., {et~al.} 2014, \apj, 783, 130

\bibitem[{{Ro{\v s}kar} {et~al.}(2008){Ro{\v s}kar}, {Debattista}, {Quinn},
  {Stinson}, \& {Wadsley}}]{ros08}
{Ro{\v s}kar}, R., {Debattista}, V.~P., {Quinn}, T.~R., {Stinson}, G.~S., \&
  {Wadsley}, J. 2008, \apjl, 684, L79

\bibitem[{{Ro{\v s}kar} {et~al.}(2012){Ro{\v s}kar}, {Debattista}, {Quinn}, \&
  {Wadsley}}]{ros12}
{Ro{\v s}kar}, R., {Debattista}, V.~P., {Quinn}, T.~R., \& {Wadsley}, J. 2012,
  \mnras, 426, 2089

\bibitem[{{Sch{\"o}nrich} \& {Binney}(2009)}]{sch09}
{Sch{\"o}nrich}, R., \& {Binney}, J. 2009, \mnras, 396, 203

\bibitem[{{Sellwood} \& {Binney}(2002)}]{sel02}
{Sellwood}, J.~A., \& {Binney}, J.~J. 2002, \mnras, 336, 785

\bibitem[{{Shapiro} {et~al.}(2010){Shapiro}, {Genzel}, \& {F{\"o}rster
  Schreiber}}]{sha10}
{Shapiro}, K.~L., {Genzel}, R., \& {F{\"o}rster Schreiber}, N.~M. 2010, \mnras,
  403, L36

\bibitem[{{Sharma} \& {Johnston}(2009)}]{sha09}
{Sharma}, S., \& {Johnston}, K.~V. 2009, \apj, 703, 1061

\bibitem[{{Steinmetz} {et~al.}(2006){Steinmetz}, {Zwitter}, {Siebert},
  {Watson}, {Freeman}, {Munari}, {Campbell}, {Williams}, {Seabroke}, {Wyse},
  {Parker}, {Bienaym{\'e}}, {Roeser}, {Gibson}, {Gilmore}, {Grebel}, {Helmi},
  {Navarro}, {Burton}, {Cass}, {Dawe}, {Fiegert}, {Hartley}, {Russell},
  {Saunders}, {Enke}, {Bailin}, {Binney}, {Bland-Hawthorn}, {Boeche}, {Dehnen},
  {Eisenstein}, {Evans}, {Fiorucci}, {Fulbright}, {Gerhard}, {Jauregi}, {Kelz},
  {Mijovi{\'c}}, {Minchev}, {Parmentier}, {Pe{\~n}arrubia}, {Quillen}, {Read},
  {Ruchti}, {Scholz}, {Siviero}, {Smith}, {Sordo}, {Veltz}, {Vidrih}, {von
  Berlepsch}, {Boyle}, \& {Schilbach}}]{ste06}
{Steinmetz}, M., {Zwitter}, T., {Siebert}, A., {et~al.} 2006, \aj, 132, 1645

\bibitem[{{Tabernero} {et~al.}(2012){Tabernero}, {Montes}, \& {Gonz{\'a}lez
  Hern{\'a}ndez}}]{tab12}
{Tabernero}, H.~M., {Montes}, D., \& {Gonz{\'a}lez Hern{\'a}ndez}, J.~I. 2012,
  \aap, 547, A13

\bibitem[{{Tabernero} {et~al.}(2014){Tabernero}, {Montes}, {Gonzalez
  Hernandez}, \& {Ammler-von Eiff}}]{tab14}
{Tabernero}, H.~M., {Montes}, D., {Gonzalez Hernandez}, J.~I., \& {Ammler-von
  Eiff}, M. 2014, ArXiv e-prints, submitted to A\&A, arXiv:1409.2348

\bibitem[{{Ting} {et~al.}(2015){Ting}, {Conroy}, \& {Goodman}}]{tin15}
{Ting}, Y.-S., {Conroy}, C., \& {Goodman}, A. 2015, \apj, 807, 104

\bibitem[{{Ting} {et~al.}(2012b){Ting}, {De Silva}, {Freeman}, \&
  {Parker}}]{tin12b}
{Ting}, Y.-S., {De Silva}, G.~M., {Freeman}, K.~C., \& {Parker}, S.~J. 2012b,
  \mnras, 427, 882

\bibitem[{{Ting} {et~al.}(2012a){Ting}, {Freeman}, {Kobayashi}, {De Silva}, \&
  {Bland-Hawthorn}}]{tin12a}
{Ting}, Y.-S., {Freeman}, K.~C., {Kobayashi}, C., {De Silva}, G.~M., \&
  {Bland-Hawthorn}, J. 2012a, \mnras, 421, 1231

\bibitem[{{Ventura} {et~al.}(2015){Ventura}, {Karakas}, {Dell'Agli}, {Boyer},
  {Garc{\'{\i}}a-Hern{\'a}ndez}, {Di Criscienzo}, \& {Schneider}}]{ven15}
{Ventura}, P., {Karakas}, A.~I., {Dell'Agli}, F., {et~al.} 2015, \mnras, 450,
  3181

\bibitem[{{Wuyts} {et~al.}(2012){Wuyts}, {F{\"o}rster Schreiber}, {Genzel},
  {Guo}, {Barro}, {Bell}, {Dekel}, {Faber}, {Ferguson}, {Giavalisco}, {Grogin},
  {Hathi}, {Huang}, {Kocevski}, {Koekemoer}, {Koo}, {Lotz}, {Lutz}, {McGrath},
  {Newman}, {Rosario}, {Saintonge}, {Tacconi}, {Weiner}, \& {van der
  Wel}}]{wuy12}
{Wuyts}, S., {F{\"o}rster Schreiber}, N.~M., {Genzel}, R., {et~al.} 2012, \apj,
  753, 114

\bibitem[{{Zasowski} {et~al.}(2013){Zasowski}, {Johnson}, {Frinchaboy},
  {Majewski}, {Nidever}, {Rocha Pinto}, {Girardi}, {Andrews}, {Chojnowski},
  {Cudworth}, {Jackson}, {Munn}, {Skrutskie}, {Beaton}, {Blake}, {Covey},
  {Deshpande}, {Epstein}, {Fabbian}, {Fleming}, {Garcia Hernandez}, {Herrero},
  {Mahadevan}, {M{\'e}sz{\'a}ros}, {Schultheis}, {Sellgren}, {Terrien}, {van
  Saders}, {Allende Prieto}, {Bizyaev}, {Burton}, {Cunha}, {da Costa},
  {Hasselquist}, {Hearty}, {Holtzman}, {Garc{\'{\i}}a P{\'e}rez}, {Maia},
  {O'Connell}, {O'Donnell}, {Pinsonneault}, {Santiago}, {Schiavon}, {Shetrone},
  {Smith}, \& {Wilson}}]{zas13}
{Zasowski}, G., {Johnson}, J.~A., {Frinchaboy}, P.~M., {et~al.} 2013, \aj, 146,
  81

\bibitem[{{Zhang} {et~al.}(2013){Zhang}, {Rix}, {van de Ven}, {Bovy}, {Liu}, \&
  {Zhao}}]{zha13}
{Zhang}, L., {Rix}, H.-W., {van de Ven}, G., {et~al.} 2013, \apj, 772, 108

\end{thebibliography}

\end{document}